\documentclass[fleqn]{2017SCGE}
\usepackage{multirow}
\usepackage{graphicx}
\usepackage{subfigure}
\usepackage{xcolor}

\begin{document}

\ensubject{subject}

\ArticleType{Article}
\SpecialTopic{SPECIAL TOPIC: }
\Year{0000}
\Month{00}
\Vol{00}
\No{0}
\DOI{xxxxxxxxxx}
\ArtNo{000000}
\ReceiveDate{00, 00, 0000}
\AcceptDate{000, 00, 0000}

\title{Influences of accretion flow and dilaton charge on the images of Einstein-Maxwell-dilation black holes }

\author[1]{Gang     Chen}{}
\author[*2]{Sen Guo}{}
\author[3]{Jia-Shuo Li}{}
\author[3]{Yu-Xiang Huang}{}
\author[*4]{Li-Fang Li}{}\thanks{Corresponding author. sguophys@126.com, lilifang@imech.ac.cn}

\author[4,5]{Peng Xu}{}




\AuthorCitation{Gang Chen, Sen Guo,Jia-Shuo Li,Yu-Xiang Huang, Li-Fang Li, Peng Xu}

\address[1]{School of Artificial Intelligence, Chongqing Technology and Business University, Chongqing 400067, China}
\address[2]{College of Physics and Electronic Engineering, Chongqing Normal University, Chongqing 401331, People's Republic of China}
\address[3]{School of Physics and Astronomy, China West Normal University, Nanchong 637000, People's Republic of China}
\address[4]{Center for Gravitational Wave Experiment, National Microgravity Laboratory, Institute of Mechanics, Chinese Academy of Sciences, Beijing 100190, China}
\address[5]{Lanzhou Center of Theoretical Physics, Lanzhou University, No. 222 South Tianshui Road, Lanzhou 730000, China.}


\abstract{The characteristics and images of Einstein-Maxwell-Dilaton (EMD) black holes are examined in this paper, focusing on their effective potential, photon trajectories, and images with thin and thick accretion disks. We found that the shadow and photon sphere radii decrease with increasing dilaton charge. As the observation inclination increases, direct and secondary images become separate, with the direct image appearing hat-shaped. Simulations indicate that the brightness of the shadow and photon ring is higher in static spherical accretion flows compared to infalling ones. The study also shows that in thin disk accretion flows, the direct emission predominantly influences observed luminosity, with photon ring emission being less significant. Additionally, the appearance of black hole images varies with the observer's inclination angle.}


\maketitle


\begin{multicols}{2}

\section{Introduction}
\label{sec:1}
\par
Black holes are among the most iconic predictions of general relativity, representing solutions to Einstein's field equations. As such, proving their actual existence has become a crucial avenue for testing the validity of Einstein's general theory of relativity. Black holes are widely regarded as celestial objects with extremely high density. Due to their immense gravitational pull, even particles traveling at the speed of light cannot escape a black hole's grasp. This means, in theory, that the black hole itself cannot be observed directly. However, on September 14, 2015, the Laser Interferometer Gravitational-Wave Observatory (LIGO) successfully detected gravitational wave signals generated by the merger of two black holes, each with a mass of 30 solar masses, thereby providing indirect evidence of their existence \cite{1}. In 2019, the Event Horizon Telescope (EHT) collaboration captured the first direct image of a black hole in the galaxy \(M87^{*}\) using a global network of radio telescopes, offering further confirmation of black holes' existence \cite{2,3,4,5,6,7}. Then, in 2022, an image of Sagittarius \(A^{*}\)(Sgr\(A^{*}\)), featuring a bright ring surrounding a dark central region, was obtained, representing the black hole's shadow. The bright ring is known as the photon ring \cite{8,9,10,11,12,13}.

\par
By studying the optical appearance of a black hole illuminated by external light sources, we can gain valuable insights into its underlying spacetime properties. Typically, black holes are surrounded by luminous accretion material, which is one of the main reasons black hole images can be observed. As light passes through the spacetime near a black hole, the intense gravitational pull causes some of the light to fall into the black hole, preventing any light from escaping. This results in the appearance of a dark disk from the observer's perspective. In 1966, Synge first examined the deflection of light around Schwarzschild black holes and proposed that the shadow of a spherical black hole would appear as a perfect circle \cite{14}. In 1973, Bardeen calculated the shadow radius of a Schwarzschild black hole as $r=3M$. For rotating black holes, the dragging effect causes the shadow to deform, producing a ``D-shaped'' silhouette \cite{15}. In 1979, Luminet created the first simulated images of the Schwarzschild black hole, employing semi-analytical methods \cite{16} to obtain both direct and secondary images of the accretion disk. Flacke et al.'s study revealed that the observed light intensity is closely correlated with the model of the accretion flow \cite{17}. Gralla et al. classified light trajectories near black holes into three categories based on the number of intersections with the equatorial plane: direct emission, lensing ring, and photon ring \cite{18}. This classification enabled a more precise description of the light rings surrounding black hole shadows.

\par
The study of the observational characteristics of black holes is an important area of research. Many researchers have explored the observational features of black holes in different gravitational spacetime backgrounds. Zeng et al. investigated the images of four-dimensional Gauss-Bonnet black holes under a spherical accretion model \cite{19}. Guo et al. examined the observational images of Hayward black holes under various accretion models. Their research suggested that the observed shadow size is closely related to the geometric properties of the black hole's spacetime, while the brightness of the shadow and ring is influenced by the accretion material and the black hole's magnetic charge \cite{20}. Additionally, numerous high-quality studies have further explored the observational characteristics of black holes in different gravitational spacetime backgrounds \cite{21,22,23,24,25,26,27,28,29,Zeng:2020vsj,Zeng:2023zlf,Zeng:2023tjb,Hu:2023mai}.

Although general relativity provides a nearly perfect description of gravity, a theory capable of simultaneously addressing both quantum effects and gravity becomes essential when attempting to understand gravity at extremely small scales. String theory aims to achieve this by describing all fundamental particles as different vibrational modes of strings. In the low-energy limit of string theory, there exists a scalar field known as the "dilaton," which is closely related to gravity and other fundamental interactions. This field can influence the curvature of spacetime, thereby altering the behavior of gravity. In this context, Gibbons and Garfinkle \cite{30,31} independently proposed extensions of static charged black hole solutions within the framework of Einstein-Maxwell-Dilaton (EMD) gravity. The action in this theory is given by:
\begin{equation}
S=\int \sqrt{-g}(R-2\nabla ^{\mu }\phi\nabla _{\mu }\phi -e^{-2\alpha \phi }F_{\mu \nu } F^{\mu \nu }  ) \mathrm{d}^{4}x,
\end{equation}
where the scalar field \(\phi\) is referred to as the dilaton field, \(R\) is the Ricci scalar, \(F_{\mu \nu}\) represents the Maxwell gauge field, and \(\alpha\) is a dimensionless parameter. The Einstein-Maxwell-Dilaton gravity theory provides a crucial theoretical framework for understanding black hole physics in the context of quantum gravity effects. Many studies have already been conducted based on this theory. For example, in \cite{36}, the trajectories of test particles and photons in the spacetime of Einstein-Maxwell-Dilaton-Axion (EMDa) black holes were studied. In \cite{37}, the shadows of EMDa black holes and naked singularities were further examined. In \cite{38}, the shadow images of charged wormholes within the EMD theory were reported, and in \cite{39}, the null geodesics and shadows of black holes in EMD gravity were discussed. When the value of parameter \(\alpha\) is fixed at 1 (\(\alpha=1\) is the value suggested by superstring theory) , we obtain the renowned Gibbons-Maeda dilaton spacetime.  There have been numerous in-depth and excellent research studies on this black hole spacetime\cite{Chen2005a,Chen2005b,Aydogdu2006,Valdez-Alvarado2017,Vieira2018,Blaga2023}.

Therefore, by studying EMD black holes, we can further explore string theory's contribution to the unification of gravity and quantum field theory, as well as investigate the role of scalar fields in curvature and gauge fields. This paper aims to study the imaging characteristics of Einstein-Maxwell-Dilaton black holes. Researching the images of EMD black holes not only provides a new perspective for understanding the interactions between the dilaton field, electromagnetic field, and gravitational field, but also aids in comprehending the low-energy manifestations of string theory and its contribution to black hole physics. Furthermore, with modern observational techniques, the study of EMD black hole images may uncover new astronomical phenomena, offering crucial experimental evidence for testing quantum gravity theories and string theory.

\par
The structure of this paper is organized as follows. In Sec.\ref{sec:2}, we analyze the effective potential of EMD black holes and utilize a ray-tracing code to plot light trajectories, along with the direct and secondary images of these black holes. In Sec.\ref{sec:3}, we simulate the images of black holes under various accretion models and examine the observational characteristics associated with each model. Finally, in Sec.\ref{sec:4}, we provide the conclusions and discussion.

\section{Effective Potential and Ray Tracing of Einstein-Maxwell-Dilaton Black Hole}
\label{sec:2}
\par
The metric of the Einstein-Maxwell-Dilaton black hole can be expressed as \cite{Vagnozzi2023,40}.
\begin{equation}
\mathrm{d}s^{2} = -f(r)\mathrm{d}t^{2} + \frac{1}{f(r)}\mathrm{d}r^{2} + r^{2}\mathrm{d}\theta^{2} + r^{2}\sin^{2}\theta \mathrm{d}\phi^{2},
\label{equ-1}
\end{equation}
Here, the metric potential \(f(r)\) of the black hole is
\begin{equation}
f(r)=1-\frac{2M}{r} \Bigg(\sqrt{1+\frac{q^{4}_{e} }{4M^{2}r^{2}}}-\frac{q^{2}_{e}}{2Mr}  \Bigg),
\label{equ-2}
\end{equation}
in which \( M \) represents the mass of the black hole and \( q_{e} \) is its dilaton charge. The relationship between the dilaton charge, electric charge, and the dilaton parameter is given by: \(q_{e}=Q\sqrt{\frac{\alpha }{8} } \) \cite{40}. When \( q_{e} \to 0 \), the EMD black hole degenerates into a Schwarzschild black hole. To investigate this black hole's images, we will study the trajectories of photons around the black hole. The motion of photons satisfies the Euler-Lagrangian equations.
\begin{equation}
\frac{\mathrm{d} }{\mathrm{d}\lambda}\Bigg(\frac{\partial\mathcal{L} }{\partial \dot{x}^{\alpha } } \Bigg)=\frac{\partial\mathcal{L} }{\partial x^{\alpha } },
\label{equ-3}
\end{equation}
here \(\lambda\) is the affine parameter and \(\dot{x}^{\alpha}\) is the four-velocity of the photon. \(\mathcal{L}\) is the Lagrangian function, given by the following equation
\begin{equation}
\mathcal{L}=-\frac{1}{2} g_{\alpha \beta }\dot{x}^{\alpha}\dot{x}^{\beta}=-\frac{1}{2}(-f(r)\dot{t}^{2}+\frac{1}{f(r)} \dot{r}^{2}+r^{2}(\dot{\theta}^{2}+\sin^{2}\theta\dot{\phi }^{2})).
\label{equ-4}
\end{equation}
For photons, \(\mathcal{L}=0\). We consider only photons moving in the equatorial plane (\(\theta = \pi/2\), \(\dot{\theta} = 0\), and \(\ddot{\theta} = 0\)). The EMD black hole metric is not a function of time \(t\) and the azimuthal angle \(\phi\), therefore two conserved quantities exist, which are
\begin{equation}
E=\Bigg(\frac{\partial \mathcal{L} }{\partial \dot{t} } \Bigg)=f(r)\frac{\mathrm{d} t}{\mathrm{d} \lambda },
\label{equ-5}
\end{equation}
\begin{equation}
L=-\Bigg(\frac{\partial \mathcal{L} }{\partial \dot{\phi } } \Bigg)=r^{2}\frac{\mathrm{d} \phi }{\mathrm{d} \lambda }.
\label{equ-6}
\end{equation}
From this, we can obtain the four-velocity components for time, azimuthal angle, and radial direction.
\begin{equation}
   \frac{\mathrm{d} t}{\mathrm{d} \lambda } =\frac{1}{b}\Bigg(1-\frac{2M}{r}\Big(\sqrt{1+\frac{q^{4}_{e} }{4M^{2}r^{2}}}-\frac{q^{2}_{e}}{2Mr}\Big)\Bigg)^{-1}
   \label{equ-7},
\end{equation}
\begin{equation}
    \frac{\mathrm{d} \phi }{\mathrm{d} \lambda } =\pm \frac{1}{r^{2}},
    \label{equ-8}
\end{equation}
\begin{equation}
    \frac{\mathrm{d} r}{\mathrm{d} \lambda } =\sqrt{\frac{1}{b^{2}}-\frac{1}{r^{2}} (1-\frac{2M}{r}(\sqrt{1+\frac{q^{4}_{e}}{4M^{2}r^{2}}}-\frac{q^{2}_{e}}{2Mr})) }.
    \label{equ-9}
\end{equation}
Here, the symbol "\(\pm\)" indicates the direction of the photon's motion on the equatorial plane (clockwise direction "\(+\)" and counterclockwise direction "\(-\)". The impact parameter \(b\) is defined as
\begin{equation}
    b=\frac{\left |  L\right | }{E} =\frac{r^{2}\dot{\phi } }{f(r)\dot{t}}.
    \label{equ-10}
\end{equation}
From Eq.(\ref{equ-9}), the effective potential of the EMD black hole is described as
\begin{equation}
    V_{eff}=\frac{1}{r^{2}} \Bigg(1-\frac{2M}{r}\Big(\sqrt{1+\frac{q^{4}_{e}}{4M^{2}r^{2}}}-\frac{q^{2}_{e}}{2Mr}\Big)\Bigg).
    \label{equ-11}
\end{equation}
The radius of the photon sphere \(r_{ph}\) and the critical impact parameter \(b_{ph}\) are obtained by the following equations 
\begin{equation}
    V(r)_{eff}=\frac{1}{b^{2}}, ~~~~~~~ V_{eff}'(r)=0.
    \label{equ-12}
\end{equation}
\begin{figure*}[htpb]
\centering
\includegraphics[scale=0.33]{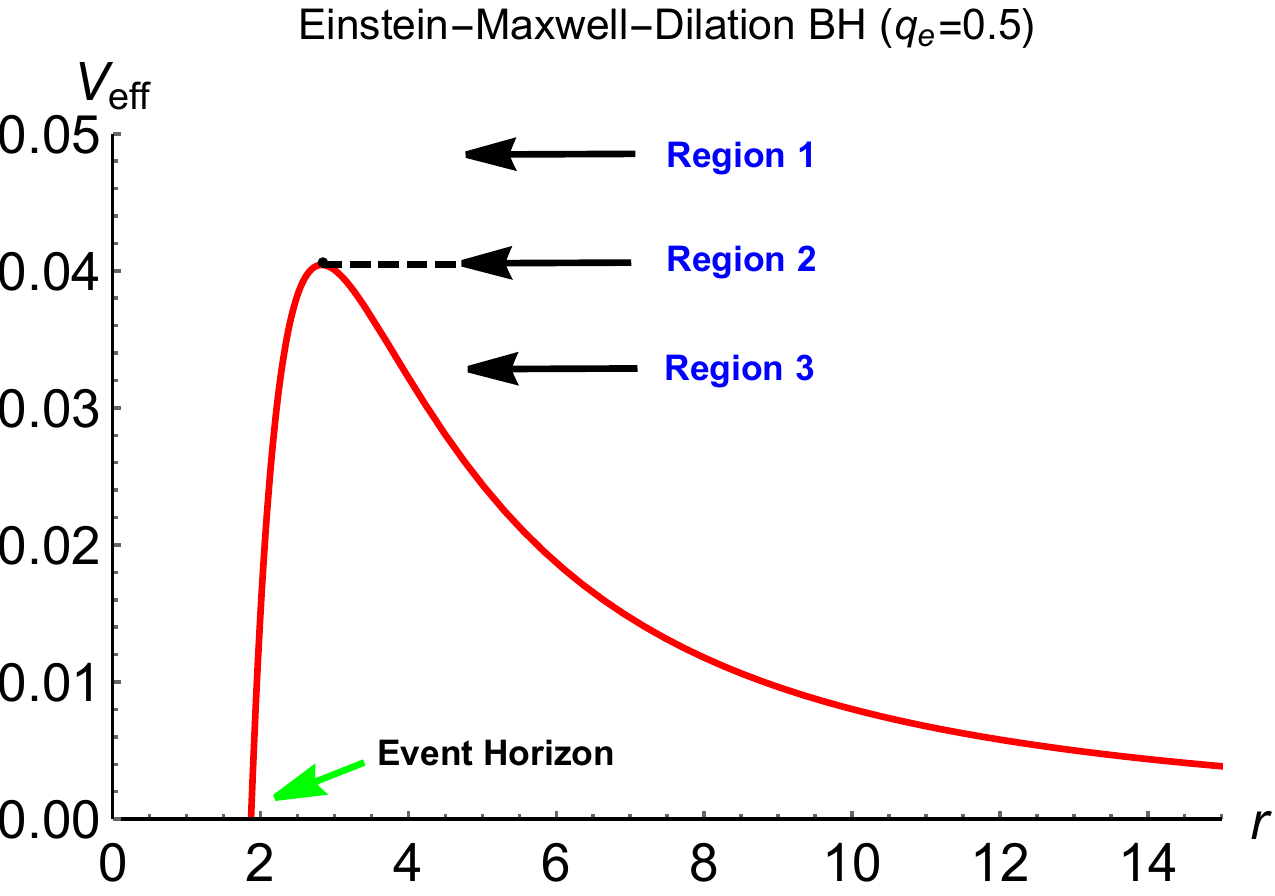}
\includegraphics[scale=0.33]{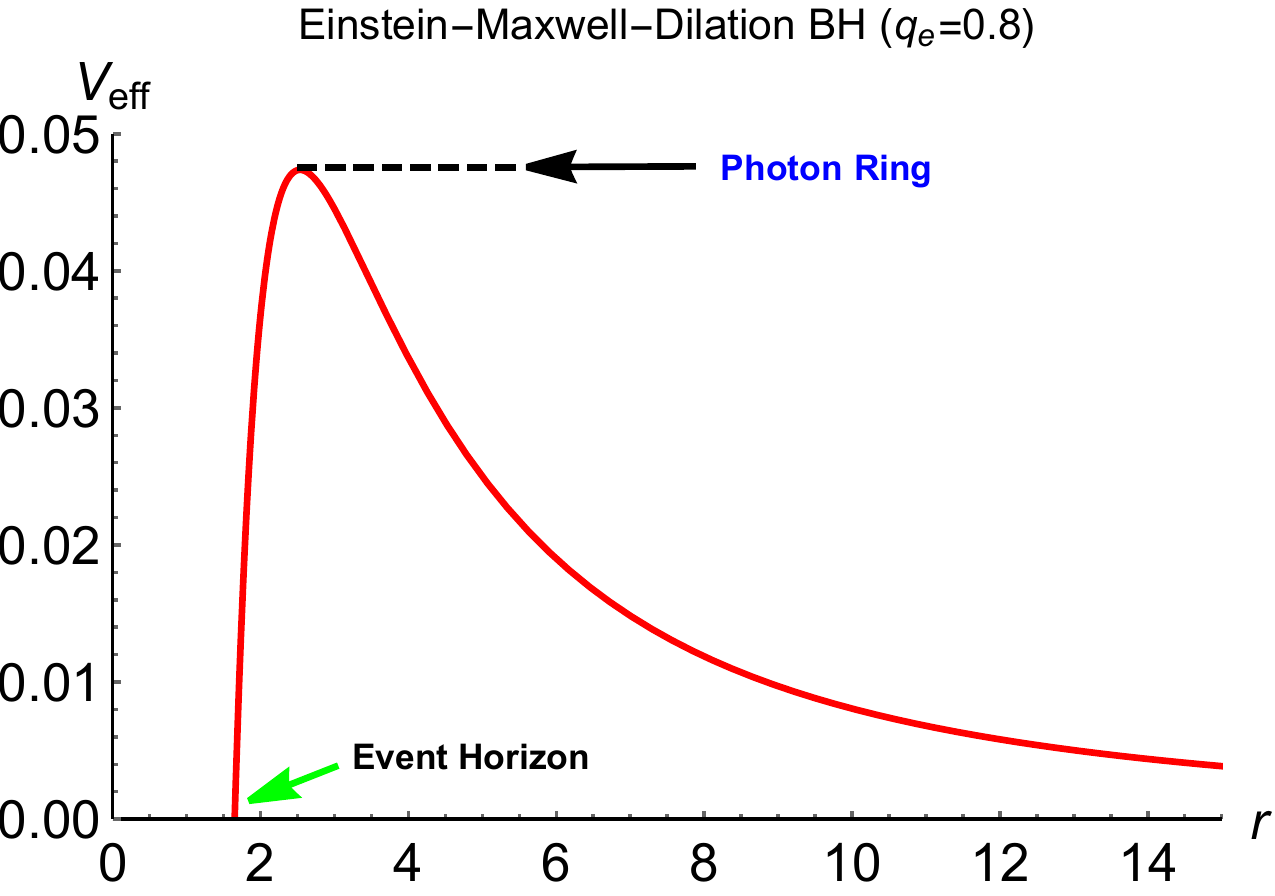}
\caption{The above figure shows the distribution of effective potential with radius. The regions labeled 1-3 partition the light rays into three areas, representing rays that fall into the black hole horizon, orbit around the black hole indefinitely, and escape to infinity, respectively. The point where the effective potential \(V_{eff}=0\) corresponds to the event horizon of the black hole. The radius at which the effective potential reaches its maximum corresponds to the photon ring radius. The left panel shows the effective potential at \(q_{e}=0.5\), and the right panel shows \(q_{e}=0.8\).}\label{fig:1}
\end{figure*}

\par
The relationship between the effective potential of the EMD black hole and the radius is shown in Figure 1. From Figure 1, it can be observed that the effective potential \(V_{eff}=0\) at the event horizon. The effective potential increases rapidly with increasing radius, reaching a maximum at the radius of the photon sphere, and then decreases as the radius continues to increase. The magnitude of the effective potential will affect the trajectory of light rays. In Region 1, where \(r < r_{ph}\), light rays will overcome the potential barrier and continue moving inward until they pass through the event horizon and enter the black hole. In Region 2, where \(r = r_{ph}\), the distance from the trajectory of the light rays to the center of the black hole equals the radius of the photon sphere; at this point, the rays will orbit the black hole along this path. In Region 3, where \(r > r_{ph}\), light rays will encounter a potential barrier, preventing them from crossing it and causing them to escape to infinity, unable to reach the black hole. From Eq.(\ref{equ-8}) and Eq.(\ref{equ-9}), we can derive
\begin{equation}
    \frac{\mathrm{d}r}{\mathrm{d}\phi } =\pm r^{2}\sqrt{\frac{1}{b^{2}}-\frac{1}{r^{2}}f(r)  }.
    \label{equ-13}
\end{equation}
Introducing a new parameter \(u \equiv 1/r\), the above equation can be rewritten as
\begin{equation}
    \Omega(u)=\frac{\mathrm{d}u}{\mathrm{d}\phi } =\sqrt{\frac{1}{b^{2}}-u^{2}\Bigg(1-2Mu\Big(\sqrt{1+\frac{u^{2}q^{4}_{e} }{4M^{2}}}-\frac{uq^{2}_{e}}{2M}\Big)\Bigg)  }.
    \label{equ-14}
\end{equation}

\begin{figure*}[htbp]
\centering
\includegraphics[scale=0.6]{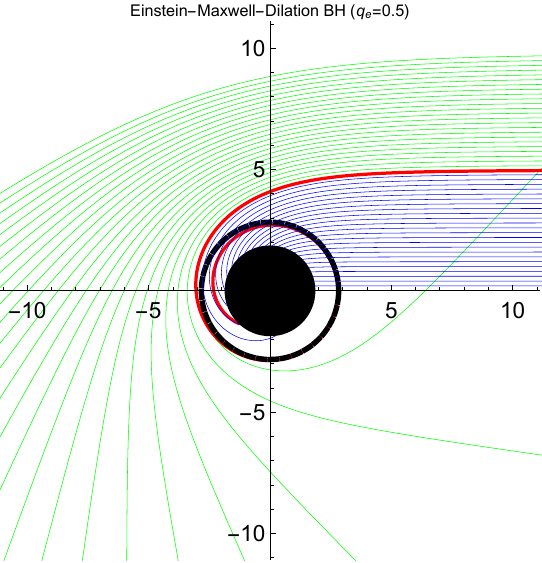}
\includegraphics[scale=0.6]{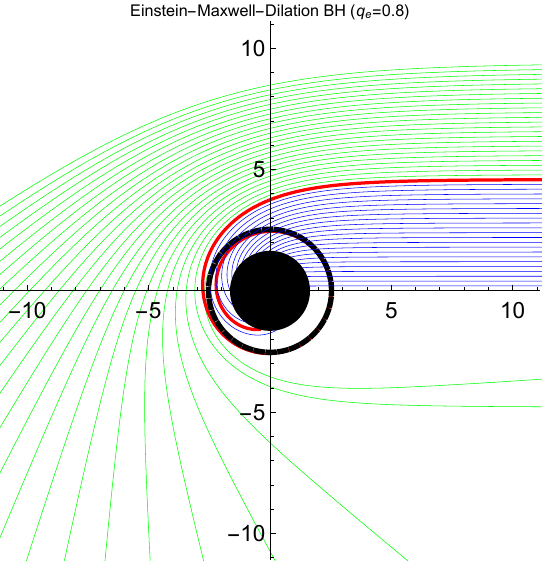}
\caption {In polar coordinates, when \(M=1\), the trajectories of light rays corresponding to different \(q_{e}\) are depicted. Blue lines represent \(b < b_{ph}\), red lines represent \(b = b_{ph}\), and green lines represent \(b > b_{ph}\). The black disk represents the black hole, and the black circle outside the disk represents the photon sphere orbit.}\label{fig:2}
\end{figure*}

\par
The deflection angle experienced by light traveling from a distant source to a faraway observer, under the influence of gravity, is expressed by the following equation:
\begin{equation}
    \psi =\int_{u_{source}}^{u^{obs}}\frac{ \mathrm{d}u }{\Omega (u)} =\int_{u_{source}}^{u^{obs}}\frac{\mathrm{d}u}{\sqrt{\frac{1}{b^{2}}-u^{2}f(u) } }.
    \label{equ-15}
\end{equation}
Using Eq.(\ref{equ-15}), light ray tracing code can be used to plot the trajectories of light rays for EMD black holes with varying dilaton charge and impact parameters. See Figure 2. In this figure, the light ray is approaching the black hole from the right side. The blue line represents \(b < b_{ph}\), the red line represents \(b = b_{ph}\), and the green line represents \(b > b_{ph}\). It can be observed that as the dilaton charge increases, both the event horizon radius and the photon sphere radius decrease. When the black hole has a larger dilaton charge, the distortion effect on light rays approaching the black hole is reduced in EMD black holes.

\par
Both EMD black holes and R-N black holes are charged black holes, each influenced by their respective charges. When both black holes have the same charge, the images of EMD black holes and R-N black holes show slight differences. For example, when the charge is \(q=0.5\), the event horizon radius of the EMD black hole is \(r_{h}=1.87083\), while the event horizon radius of the R-N black hole is \(r_{h}=1.86603\). From Figure 3, we can observe that the EMD black hole causes a stronger distortion of light compared to the R-N black hole. Therefore, theoretically, we can distinguish between the two black holes based on their images.
\begin{figure}[H]
\centering
\includegraphics[scale=0.6]{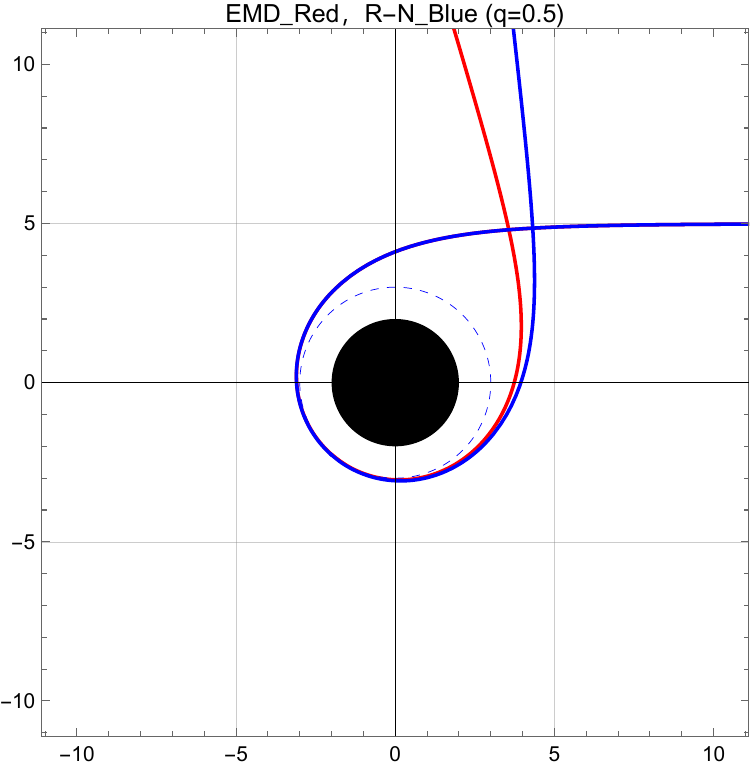}
\caption {The image shows the trajectory of the same light ray after being distorted by the EMD black hole and the RN black hole. The red line represents the distortion caused by the RN black hole, while the blue line represents the distortion caused by the EMD black hole.}\label{fig:3}
\end{figure}

\par
To vividly illustrate the distortion of spacetime by the black hole, the background light source is divided into four quadrants, each painted with a different color. In Figure 4, the black grid lines represent longitude and latitude. This image is obtained through inverse ray tracing from the observer to the background light source. The black disk at the center of the image represents the shadow of the black hole, while the bright and colorful patterns depict the background light source distorted by the black hole. The image features surrounding the black disk are a result of the gravitational lensing effect of the black hole. 
\begin{figure}[H]
\centering
\includegraphics[scale=0.55]{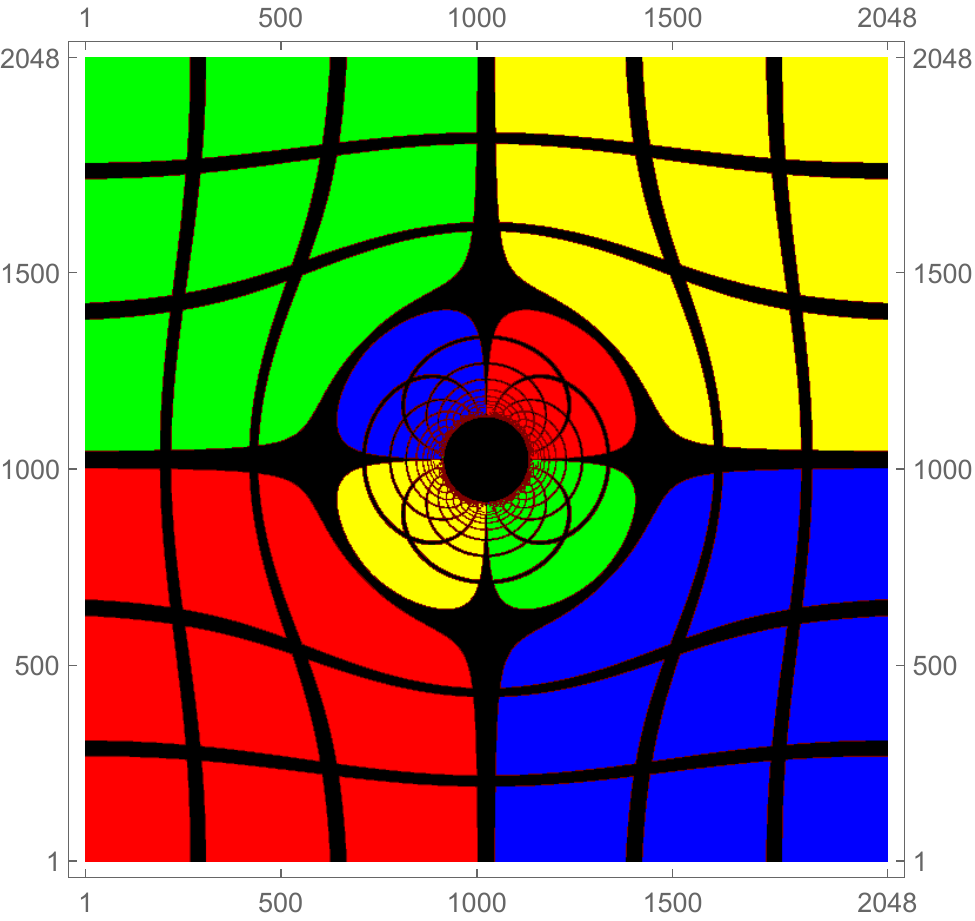}
\caption {The image displays the distorted trajectories caused by the gravitational lensing effect of the black hole, taking into account the background source at infinity.}\label{fig:4}
\end{figure}

\par
Assuming the radiation is emitted from the point \((r, \phi)\) on the emission plane \(M\), the observer can detect the light at the point \((b, \alpha)\) on the observation plane \(m\). For the light rays emitted from the emitting plane, the direct image with coordinates \((b^{(d)},\alpha)\) will be observed on the observer plane, and the secondary image with coordinates \((b^{(s)},\alpha)\). Assuming the azimuthal angle of the photons emitted by the radiation source is \(\phi\), there exists a relationship with the celestial coordinates \(\eta\) and the observer's inclination angle \(\theta_{0}\).
\begin{equation}
    \cos \phi =-\frac{\sin\eta\ast \tan \theta _{0}}{\sqrt{\sin^{2}\eta\ast\tan^{2}\theta _{0}+1    } },
    \label{equ-16}
\end{equation}
\begin{equation}
    \sin \phi =-\frac{1}{\sqrt{\sin^{2}\eta\ast\tan^{2}\theta _{0}+1    } }.
    \label{equ-17}
\end{equation}
Through Eqs.(\ref{equ-16}-\ref{equ-17}), the deflection angle experienced by the light traveling from the emission source to the distant observer can be obtained as
\begin{equation}
   \phi= \int_{u_{source}}^{u_{o b s}} \frac{d u}{\sqrt{\frac{1}{b^{2}}-u^{2} f(u)}}=-\arccos \frac{\sin \eta \ast\tan \theta_{0}}{\sqrt{\sin ^{2} \eta\ast \tan ^{2} \theta_{0}+1}}.
    \label{equ-18}
\end{equation}
Due to the strong gravitational field around the black hole, some light rays may undergo multiple deflections around the black hole. The above equation may no longer be applicable in these cases, requiring consideration of more general scenarios.
\begin{equation}
\int_{u_{source }}^{u_{obs}} \frac{d u}{\sqrt{\frac{1}{b^{2}}-u^{2} f(u)}}=k \pi-\arccos \frac{\sin \eta \ast\tan \theta_{0}}{\sqrt{\sin ^{2} \eta\ast \tan ^{2} \theta_{0}+1}}
    \label{equ-19},
\end{equation}
where \( k \) is an integer representing the order of the image: \( k=0 \) denotes the direct image of the black hole accretion disk, \( k=1 \) denotes the secondary image, and \( k>1 \) indicates higher-order images. Therefore, the direct and secondary images at equal radii of the accretion disk around the black hole can be plotted using Eq.(\ref{equ-19}).

\end{multicols}
\begin{figure*}[htbp]
  \centering
  \includegraphics[width=5cm,height=5cm]{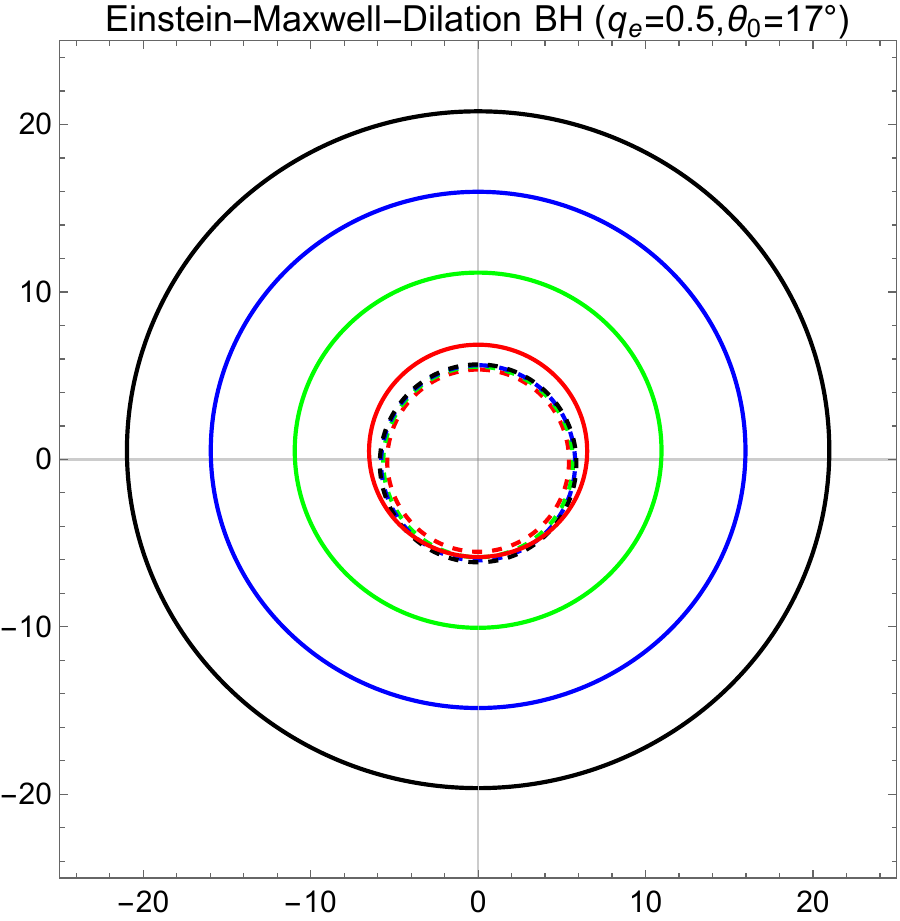}
  \includegraphics[width=5cm,height=5cm]{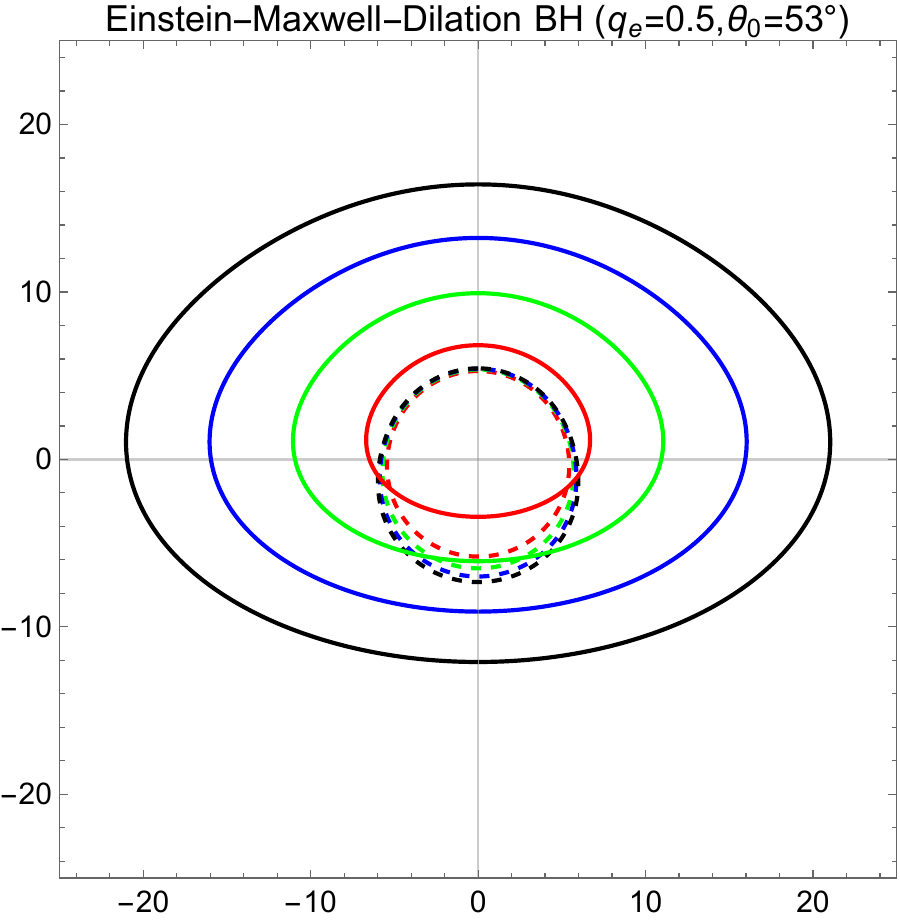}
  \includegraphics[width=5cm,height=5cm]{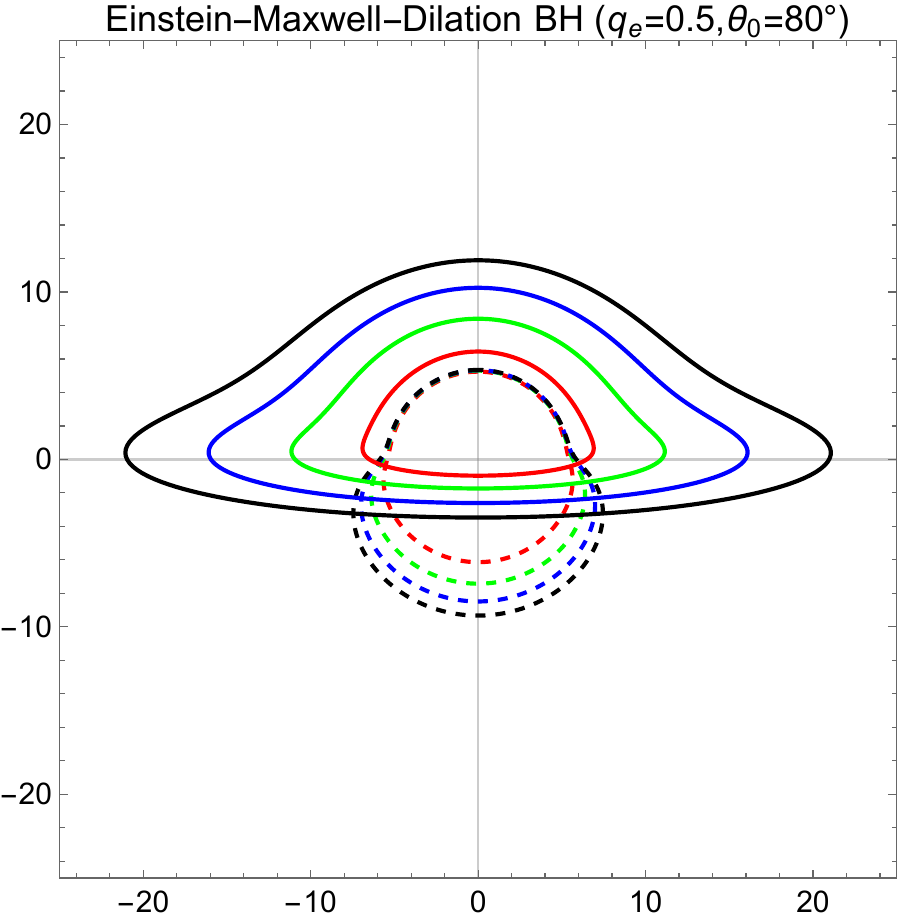}
  \includegraphics[width=5cm,height=5cm]{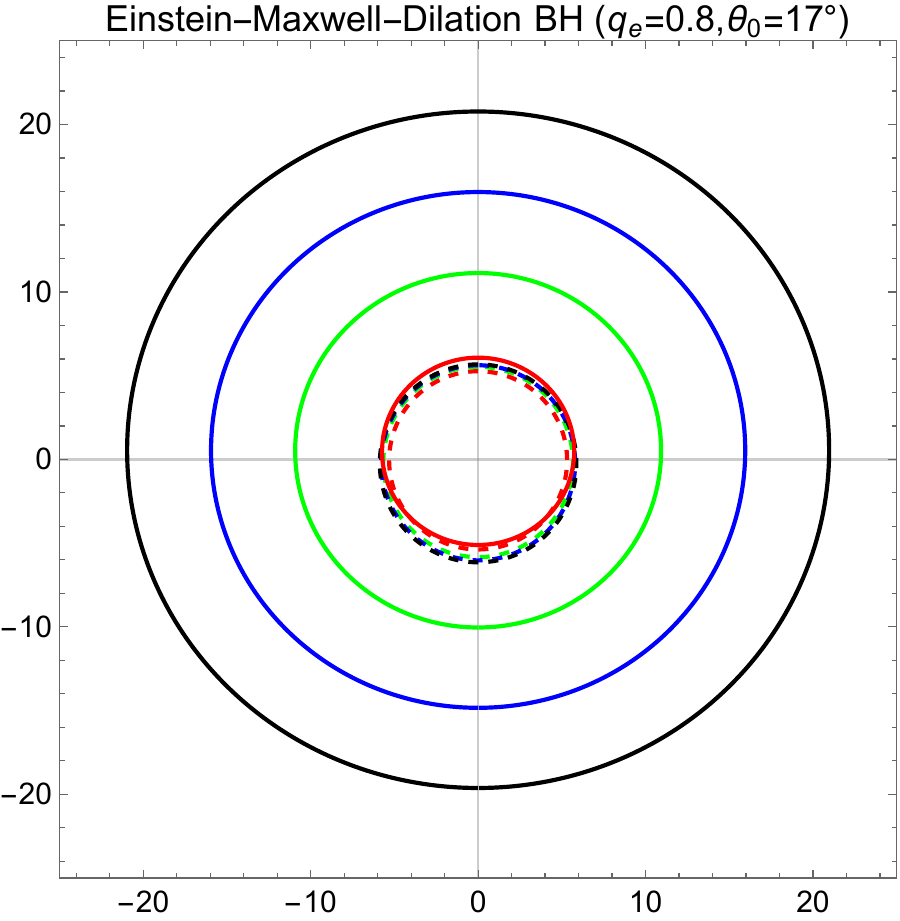}
  \includegraphics[width=5cm,height=5cm]{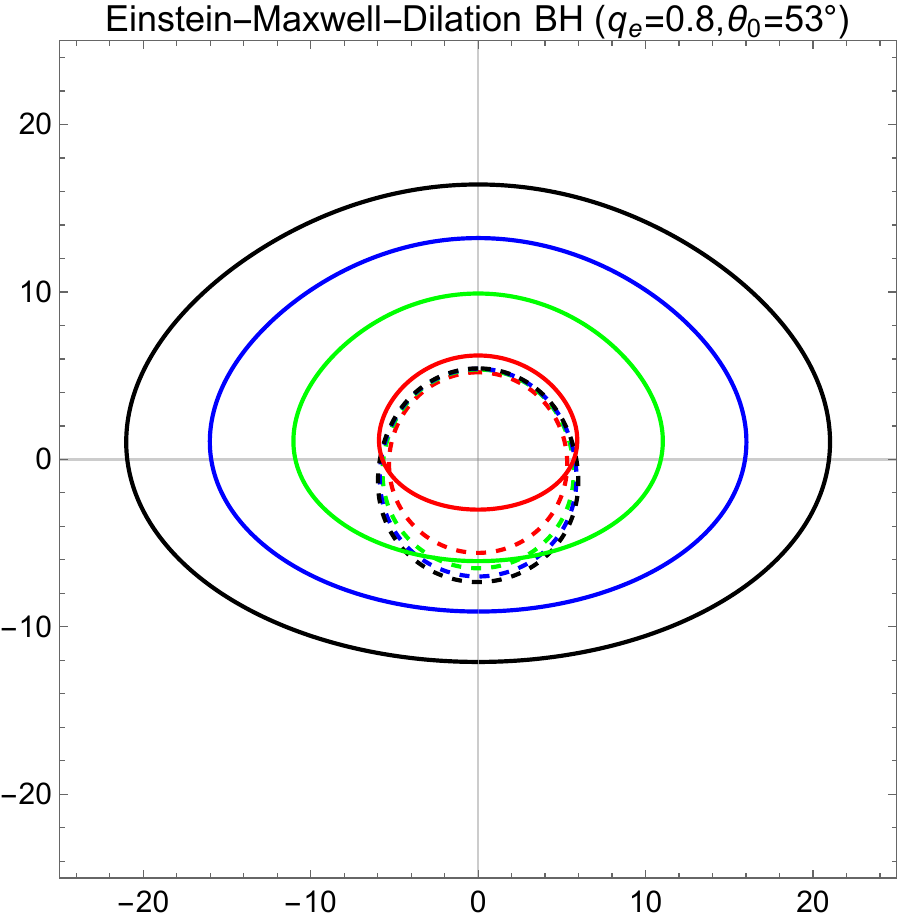}
  \includegraphics[width=5cm,height=5cm]{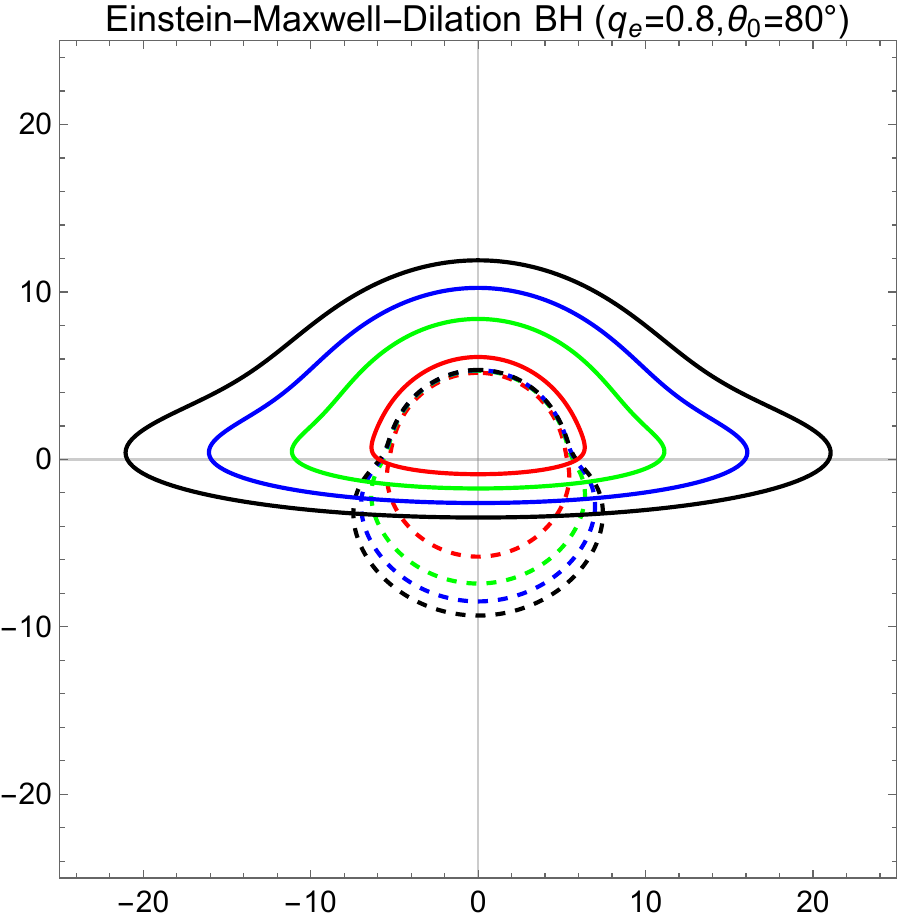}
  \caption {Images of circular orbits at various radii around a black hole observed from three different inclinations. Different colors in the image represent different radii: red, green, blue, and black correspond to radii of 6M, 10M, 15M, and 20M, respectively. }The first row corresponds to an EMD black hole with \(q_{e}=0.5\). The second row corresponds to an EMD black hole with \(q_{e}=0.8\).\label{fig:5}
\end{figure*}
\begin{multicols}{2}

\par
Figure 5 shows the direct and secondary images of circular orbits around the black hole at different observation angles. The solid lines represent direct images, while the dashed lines correspond to the secondary images of the same color. It can be observed that as the observation angle increases, the secondary images of the ring structure initially embedded within the direct images gradually separate from the direct images. Additionally, the direct images transform from a ring shape to a hat shape. The upper and lower rows in the image represent the direct and secondary images of black holes with different dilaton charges, respectively. It can be observed that as the dilaton charge increases or decreases, the gap between the direct images and the secondary images becomes smaller.

\par

When \(k=2\), higher-order images will appear. These higher-order images are extremely narrow and exist at the inner edge between the direct image and the secondary image, forming a thin ring-like structure. To observe such narrow light rings, extremely high resolution is required for detection. Their contribution to the black hole image is very weak. As shown in Figure 6, the higher-order images are represented by the ring formed by the dotted lines within the secondary image. It can be observed that the images at radii of 6M, 10M, 15M, and 20M overlap, meaning that the higher-order images will form an extremely narrow ring, which is difficult to distinguish in the image.

\begin{figure}[H]
\centering
\includegraphics[scale=0.55]{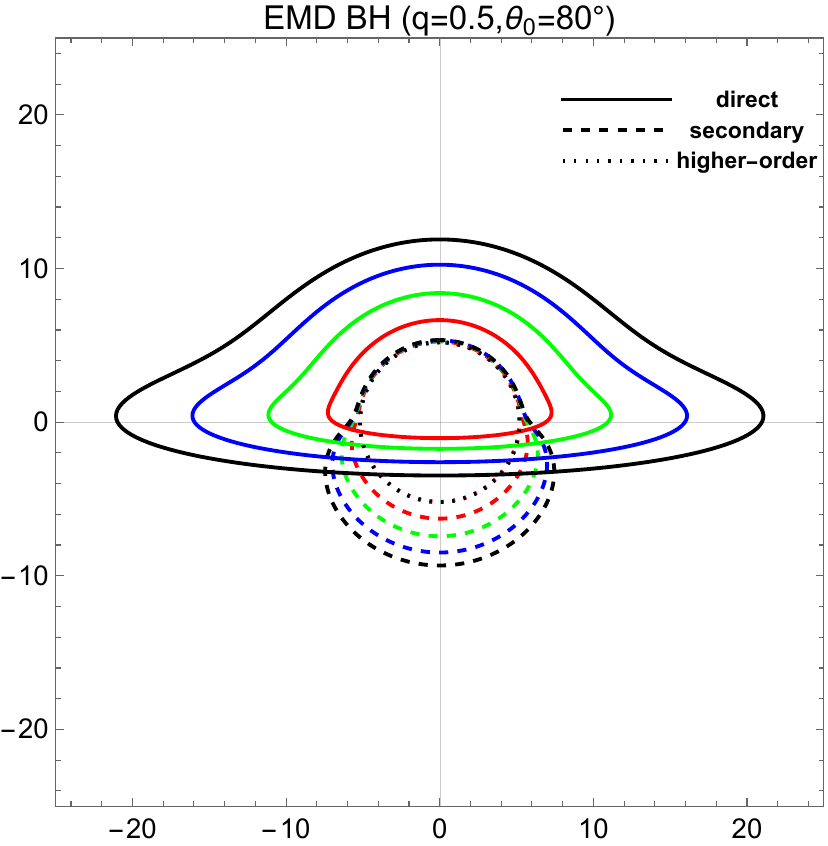}
\caption {The image represents the direct, secondary, and higher-order images of the black hole when \(q=0.5\) and \(\theta = 80^{\circ} \). The solid lines indicate the direct images, the dashed lines represent the secondary images, and the dotted lines correspond to the higher-order images.} \label{fig:6}
\end{figure}

\section{Images of black holes with different accretion models}
\label{sec:3}
\subsection{Static spherically symmetric accretion model}
\label{sec:3-1}
\par
This section will study the observational characteristics of a black hole surrounded by a static, optically thin, and geometrically thin spherical accretion flow. First, we need to determine the specific intensity observed by the observer. The intensity of light observed at the frequency \( v_{obs}^{s} \) can be expressed as follows.
\begin{equation}
    I(v_{obs}^{s} )=\int g^{s3} j(v_{em}^{s})\mathrm{d}l_{prop},
    \label{equ-20}
\end{equation}
where \(g^{s}\equiv \frac{v_{obs}^{s}}{v_{em}^{s}}\) is the redshift factor, \(v_{em}^{s}\) is the intrinsic frequency of the photon, \(\mathrm{d}l_{prop}\) is the infinitesimal proper length, and \(j(v_{em}^{s})\) is the emission per unit volume in the rest frame of the emitting material. For a four-dimensional spherically symmetric black hole, the redshift factor can be expressed as \(g^{s} \equiv f(r)^{1/2}\). Here, we consider a simple case where the emission is monochromatic with a frequency \(v_{t}\) in the rest frame, and the radial distribution of the emission is \(1/r^{2}\), that is 
\begin{equation}
    j(v_{em}^{s})\propto \frac{\delta (v_{em}^{s}-v_{t})}{r^{2}}.
    \label{equ-21}
\end{equation}
According to Eq.(\ref{equ-1}), the proper length in the rest frame of the emitting material is given by
\begin{equation}
\mathrm{d}l_{prop}=\sqrt{f(r)^{-1}\mathrm{d}r^{2}+r^{2}\mathrm{d}\phi^{2}} =\sqrt{f(r)^{-1}+r^{2}\Bigg(\frac{\mathrm{d} \phi }{\mathrm{d} r} \Bigg)^{2}}\mathrm{d}r.
    \label{equ-22}
\end{equation}
Using Eqs.(\ref{equ-20}-\ref{equ-22}), the total photon intensity measured by a distant observer can be derived as
\begin{equation}
  I(v_{obs}^{s} )=\int \frac{f(r)^{\frac{3}{2} }}{r^{2}}\sqrt{\frac{1}{f(r)}+\frac{b^{2}r^{4}}{r^{2}-f(r)b^{2}}  }  \mathrm{d}r
  \label{equ-23}.
\end{equation}
Next, Eq.(\ref{equ-23}) will be used to study the shadow of a black hole with a static spherical accretion model.

\par
Figure 7 shows the observed intensity \(I(v_{obs}^{s} )\) as a function of the impact parameter \(b\) for different values of \(q_{e}\). The observed intensity increases with the impact parameter, reaching a peak at \(b=b_{ph}\), and then decreases as the impact parameter continues to increase. Additionally, for relatively large \(q_{e}\), the intensity is also higher. When \(b<b_{ph}\), due to the proximity of the accreting material to the black hole, most of the generated light is absorbed by the black hole. For an optically thin source, the intensity accumulates along the path, so the path length of the light represents its intensity. When \(b = b_{ph}\), the light rays orbit the black hole multiple times, resulting in maximum intensity at this point. For \(b > b_{ph}\), only refracted light contributes to the observed intensity. As the impact parameter \(b\) increases, the refracted light gradually decreases, leading to a reduction in the observed intensity.
\begin{figure}[H]
\centering
\includegraphics[scale=0.55]{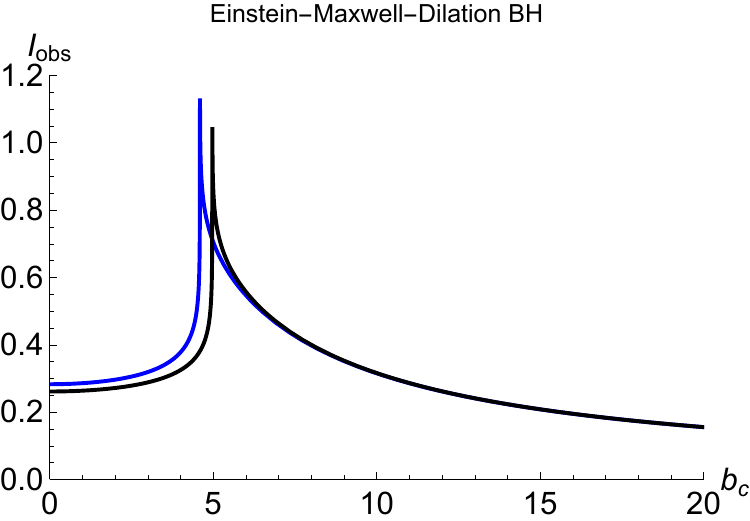}
\caption {The relationship between the photon intensity \(I(v_{obs}^{i})\) and the impact parameter in the static spherical accretion flow model. The black line represents the intensity measured by the observer for \(q_{e}=0.5\), while the blue line represents the intensity for \(q_{e}=0.8\). In both cases, the intensity increases with the impact parameter \(b\) until it reaches a maximum at \(b=b_{ph}\), after which it decreases as \(b\) continues to increase.}\label{fig:7}
\end{figure}

\par
Figure 8 shows the two-dimensional projection of the shadow of an EMD black hole with static spherical accretion flow in celestial coordinates. From the figure, it can be seen that there is a bright circular ring surrounding a central darkened region, which is known as the black hole's shadow. The shadow of a black hole is not a completely dark region with zero light intensity in that some radiation from the accretion flow within the photon sphere can escape to infinity. We can also observe that for the brightness of the EMD black hole shadow and the photon ring, the black hole with a smaller dilaton charge \(q_{e}\) is dimmer compared to the black hole with a larger \(q_{e}\).
\begin{figure*}[htbp]
\centering
\includegraphics[scale=0.45]{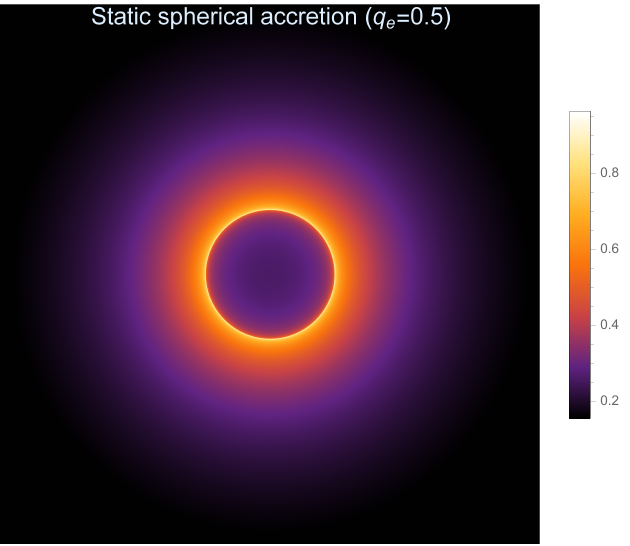}
\includegraphics[scale=0.45]{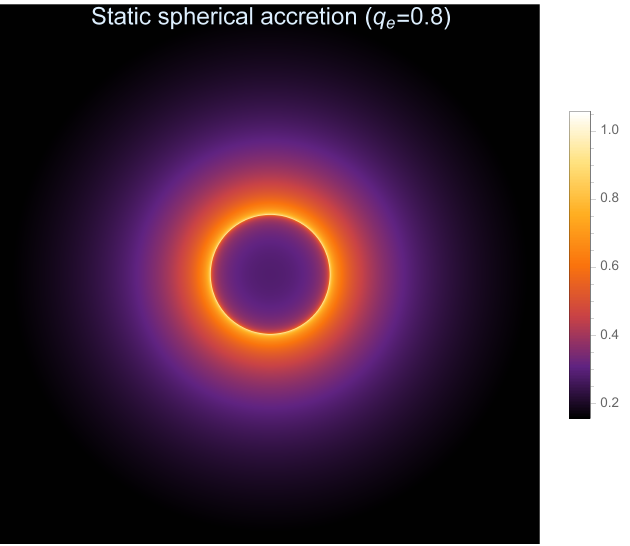}
\caption {The two-dimensional images of the shadow and photon ring for EMD black hole shadow with \(q_{e}=0.5\) and \(q_{e}=0.8\) under the static spherical accretion model with \(M=1\).}\label{fig:6}
\end{figure*}

\subsection{\textbf{Infalling spherically symmetric accretion model}}
\label{sec:3-2}
\par
Since most accretion in the real universe involves matter falling into black holes, in this section, we will consider an optically thin spherical accretion model where matter falls into the black hole. This model is more realistic than the static spherical accretion. In this scenario, the intensity observed by the observer, governed by Eq.(\ref{equ-20}), remains effective. However, due to matter falling into the black hole, the redshift factor will change as 
\begin{equation}
   g^{i}=\frac{k_{\rho }u_{obs}^{i\rho}}{k_{\sigma}u_{em}^{i\sigma}}
  \label{equ-24},
\end{equation}
where \(k_{\mu}\) is the four-velocity of the photon, \(u_{obs}^{i\mu}=(1,0,0,0)\) is the four-velocity of the distant observer, and \(u_{em}^{i\mu}\) is the four-velocity of the accretion flow. According to Eqs.(\ref{equ-1}-\ref{equ-9}), \(k_{t}\) is a constant, i.e., \(k_{t} \equiv 1/b\). Meanwhile, \(k_{r}\) can be derived from \(k_{\gamma}k^{\gamma}=0\) and thus, we have
\begin{equation}
 \frac{k_{r}}{k_{t}}=\pm\sqrt{\frac{1}{f(r)}\Bigg(\frac{1}{f(r)}-\frac{b^{2}}{r^{2}}\Bigg)},
    \label{equ_25}
\end{equation}
here $\pm$ indicate whether the photon is approaching or receding from the black hole. For the four-velocity of the accretion flow, we have
\begin{equation}
    u_{em}^{it}=\frac{1}{f(r)},
    \label{equ_26}
\end{equation}
\begin{equation}
   u_{em}^{ir}=-\sqrt{1-f(r)},
    \label{equ_27}
\end{equation}
\begin{equation}
  u_{em}^{i\theta }=u_{em}^{i\phi}=0.
    \label{equ_28}
\end{equation}
From Eqs.(\ref{equ_26}-\ref{equ_28}), the redshift factor is written as
\begin{equation}
    g=\frac{1}{u_{em}^{it}+(\frac{k_{r}}{k_{em}^{t}} )u_{em}^{ir}}.
    \label{equ-29}
\end{equation}
The infinitesimal proper length is given by
\begin{equation}
    \mathrm{d}l_{prop}=k_{\sigma}u_{em}^{i\sigma}\mathrm{d} \lambda =\frac{k_{t}}{g^{i3}\left |  k_{r}\right | } \mathrm{d}r.
    \label{equ_30}
\end{equation}
Thus, the observed intensity in the infalling spherical accretion model is given by
\begin{equation}
    I(v_{obs}^{i})=\int \frac{g^{i3}}{r^{2}\sqrt{\frac{1}{f(r)}(\frac{1}{f(r)}-\frac{b^{2}}{r^{}}  ) } }\mathrm{d}r.
    \label{equ_31}
\end{equation}
\par
Figure 9 shows the relationship between the total observed intensity and the impact parameter. The overall trend of the function is similar to that in Figure 7, but the total observed intensity decreases more rapidly for \(b > b_{ph}\), approaching zero. For the accretion model of infalling spherical matter, its overall total observed intensity is lower than that of the stationary spherical accretion model. Figure 10 shows the two-dimensional projection of this model in observer's celestial coordinates. From the image, it can be seen that this model exhibits the same position and size of the black hole shadow in the static spherical accretion model. Due to the more rapid decline in observed intensity in this model, even eventually approaching zero, the radius of the bright region in the observed image is smaller, and the overall intensity of the observed image is significantly reduced.
\begin{figure}[H]
\centering
\includegraphics[scale=0.55]{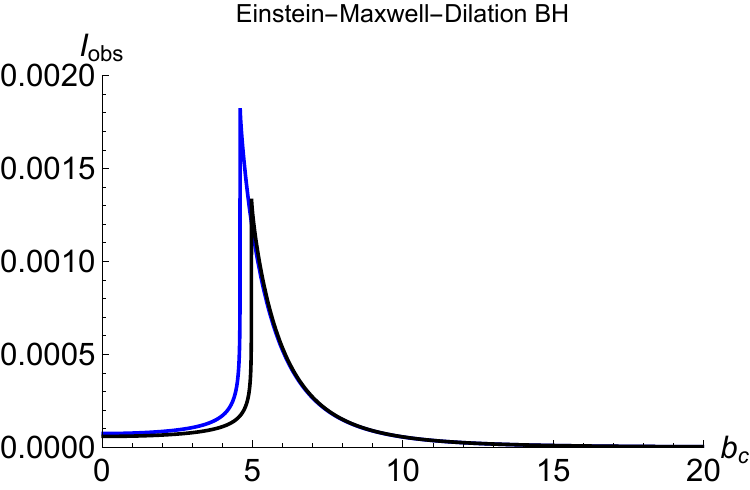}
\caption {For the infalling spherical accretion flow model, the photon intensity \(I(v_{obs}^{i})\) as a function of the impact parameter \(b\) is shown in the figure. The black line represents the observed intensity for \(q_{e}=0.5\). The blue line represents the observed intensity for \(q_{e}=0.8\). The intensity increases with increasing impact parameter \(b\) until \(b=b_{ph}\), where it reaches its maximum, and then decreases as \(b\) further increases.}\label{fig:9}
\end{figure}
\begin{figure*}[htbp]
\centering
\includegraphics[scale=0.45]{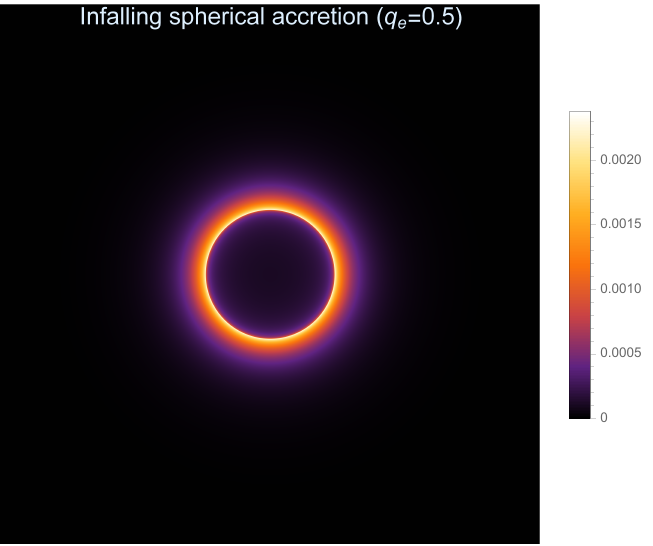}
\includegraphics[scale=0.45]{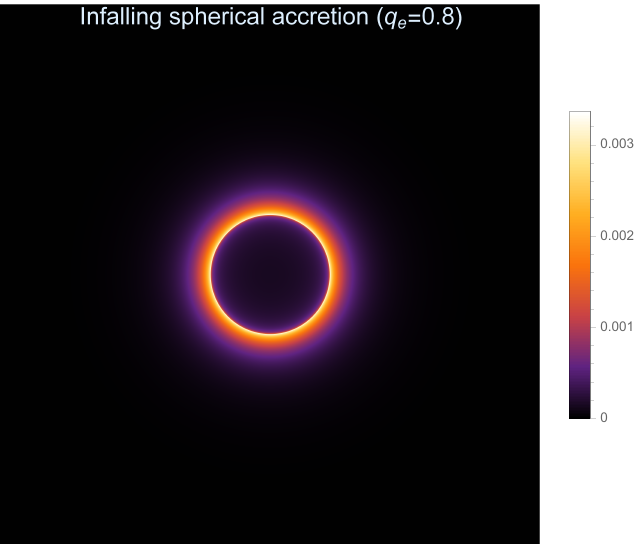}
 \caption {The two-dimensional image of the shadow and photon ring of EMD black hole with \(M=1\) under the infalling spherical accretion flow model, for \(q_{e}=0.5\) and \(q_{e}=0.8\).}\label{fig:10}
\end{figure*}

\subsection{\textbf{Thin disk accretion model}}
\label{sec:3-3}
\par
In this section, we will consider two aspects of the black hole accretion disk model: the geometric thickness and the optical thickness of the accretion disk. The geometric thickness of the accretion disk refers to the scale in the direction perpendicular to the radial direction of the disk and is typically quantified by the ratio of the disk's thickness \(H\) to its radius \(R\) (i.e., \(H/R\)). If \(H/R\ll 1\), the accretion disk is referred to as a "thin disk". Optical thickness describes the transparency of the accretion disk to photon propagation. If photons find it difficult to pass through the accretion disk, the disk is considered optically thick.
In this model, it is assumed that there exists an optically thin geometrically thin accretion disk in the equatorial plane of the black hole, with the observer located at the north pole of the black hole. According to the definition in \cite{18}, the total number of light ray trajectories is (\(n\equiv\phi/2\pi\)). Additionally, light rays can be classified into three types:
\begin{enumerate}
    \item Direct emission (\(n<3/4\)): Light ray trajectories intersect the accretion disk in the equatorial plane only once.
    \item Lensed ring (\(3/4<n<5/4\)): Light ray trajectories intersect the accretion disk in the equatorial plane twice.
    \item Photon ring (\(n>5/4\)): Light ray trajectories intersect the accretion disk in the equatorial plane three times or more.
\end{enumerate}

\par
Figure 11 illustrates the relationship between the number of orbits and the collision parameter. In the figure, the red region represents direct emission, the green region represents lensed emission, and the blue region represents photon ring emission. From the image, it can be observed that for a black hole with \(q_{e}=0.5\) the collision parameters corresponding to the photon ring and lensed emission are larger compared to a black hole with \(q_{e}=0.8\). Figure 12 depicts photon trajectories plotted using the ray tracing code, where regions are defined in colors consistent with the previous descriptions. The first image shows the trajectory for \( q_e = 0.5 \), while the second image shows the trajectory for \(q_e=0.8\). It is clear from these two images that an increase in dilaton charge causes the photon ring to thicken.
\begin{figure}[H]
\centering
\includegraphics[scale=0.55]{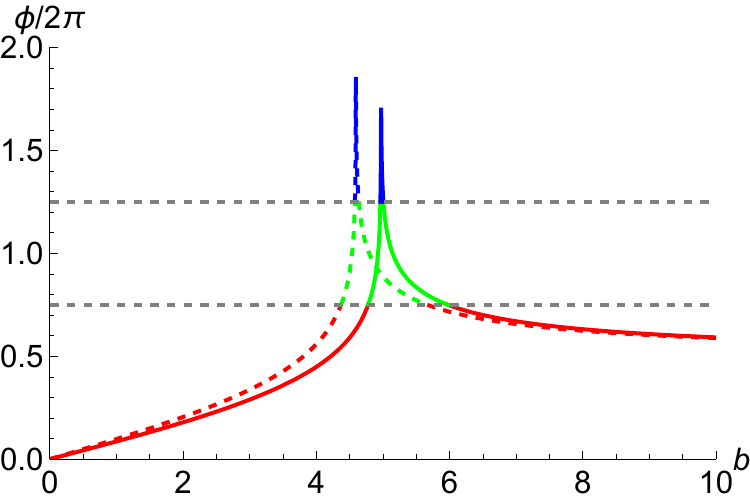}
 \caption {The relationship between the number of orbits and the collision parameter is shown in the figure. The red region represents direct emission, the green region represents lensed emission, and the blue region represents photon ring emission. The solid line corresponds to a black hole with \( q_e = 0.5 \), while the dashed line corresponds to a black hole with \( q_e = 0.8 \).}\label{fig:11}
\end{figure}
\begin{figure*}[htbp]
\centering
\includegraphics[scale=0.38]{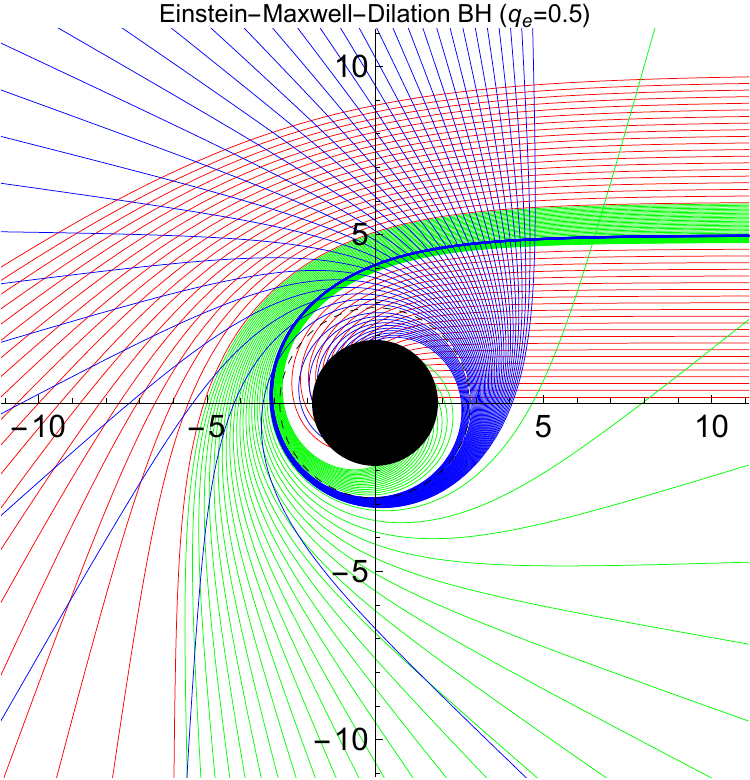}
\includegraphics[scale=0.38]{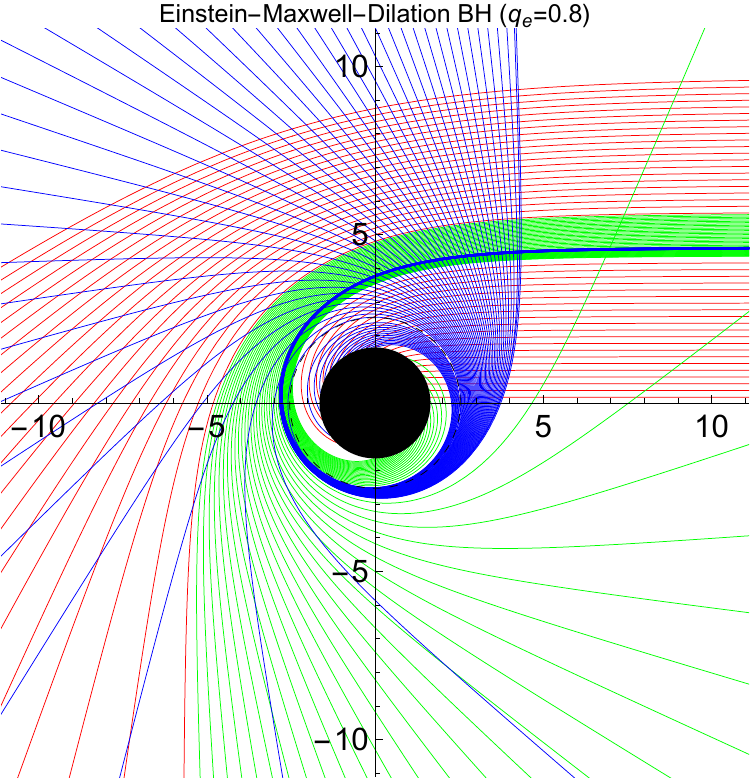}
 \caption {Photon trajectory diagrams of EMD black holes. In these diagrams, red, blue, and green denote direct emission, lensed emission, and photon ring, respectively. The left panel corresponds to an EMD black hole with \( q_e = 0.5 \), while the right panel corresponds to an EMD black hole with \( q_e = 0.8 \).}\label{fig:12}
\end{figure*}

\par
Since a thin accretion disk is optically thin, each time a light ray passes through it, additional brightness is added from the disk. Therefore, as the number of intersections with the disk increases, the intensity of the light ray also increases. Therefore, the observed total intensity should be the sum of the intensities at each intersection point. The study of specific intensities for a thin accretion disk considers the emission from the disk to be isotropic. According to Liouville's theorem, \( I_{em}/(v_{em}^{d})^{3} \) is conserved along the direction of light propagation, hence \( I_{em}/(v_{em}^{d})^{3} = I_{obs}/(v_{obs}^{d})^{3} \), where \( I_{em} \) is the emitted specific intensity and \( v_{em}^{d} \) is the emitted frequency. The relationship between the observed specific intensity at a single frequency and the emitted specific intensity is given by
\begin{equation}
    I_{obs}^{d}(r)=g^{3}I_{em}^{d}=f(r)^{\frac{3}{2}}I_{em}^{d}.
    \label{equ_32}
\end{equation}
The total observed intensity is obtained by integrating Eq.(\ref{equ_32}) over the entire observed frequency range
\begin{equation}
    I_{o}=\int I_{obs}^{d}(r)\mathrm{d}v_{obs}^{d} =\int f(r)^{2}I_{em}^{d}\mathrm{d}v_{em}^{d}=f(r)^{2}I_{em}^{d},
    \label{equ_33}
\end{equation}
where \( I_{emi}^{d} \equiv \int I_{em}^{d} \mathrm{d}v_{em}^{d} \). The total observed intensity is given by
\begin{equation}
    I_{o}=\sum_{n}f(r)^{2}I_{em}^{d}(r)\mid _{r=r_{n}(b)},
    \label{equ_34}
\end{equation}
in which \( r_{n}(b) \) is the transfer function representing the radial position of the \(n\)-th intersection of a light ray with the thin disk at collision parameter \(b\). 

\par
Figure 13 displays the first three transfer function plots for black holes with \( q_{e} = 0.5 \) and \( q_{e} = 0.8 \). The slope \( d(r)/d(b) \) of the transfer function acts as a magnification factor, indicating the degree of size reduction in the observed image. The red curve in the figure represents \( n=1 \), where the slope of the transfer function is \( 1 \), indicating the direct image from the redshift source. The green curve represents \( n=2 \), where the slope of the transfer function is much greater than \( 1 \), corresponding to an image where the disk's far side is highly magnified, corresponding to the lensed ring. The blue curve represents \( n=3 \), where the slope tends to infinity, resulting in an image where the disk's near side is greatly reduced, making the contribution of the photon ring to the total intensity negligible. From the figures, it is also evident that for EMD black holes with different dilaton charges \( q_{e} \), as the dilaton charge increases, the collision parameters corresponding to these images decrease.
\begin{figure}[H]
\centering
\includegraphics[scale=0.55]{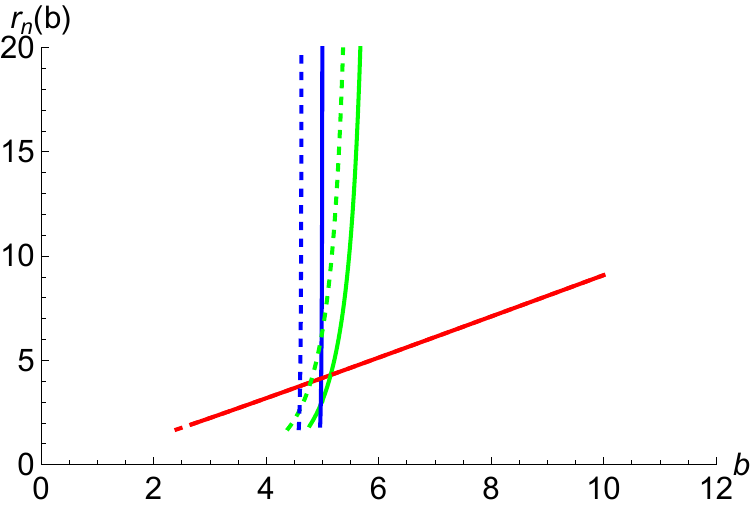}
\caption {Transfer functions for different values of \( q_{e} \), where the solid line corresponds to a black hole with \( q_{e} = 0.5 \), and the dashed line corresponds to a black hole with \( q_{e} = 0.8 \). The red line represents direct emission, the green line represents the lensed ring, and the blue line represents the photon ring.}\label{fig:13}
\end{figure}

\par
Next, we will investigate the influence of different emission locations in the accretion disk on the observational characteristics of black holes. Given that we aim to compare images under different dilaton charges, the accretion disk emission model we adopt is merely a toy model. These models simplify the physical processes of the accretion disk, disregarding the complex magnetohydrodynamic behavior of the plasma. The focus is primarily on the simple geometry and radiation behavior of the accretion disk, without considering the complex structure of the disk or the direct influence of the surrounding environment on the radiation. Existing research has indicated that the radiation from accretion disks follows a Gaussian distribution\cite{S.N.Zhang:2013}. Consequently, the radiation from the accretion disk is parameterized as a Gaussian function in this context.
\begin{equation}
    I_{\mathrm{emi}}^{\mathrm{d}}(r)=\left\{\begin{array}{ll}e^{\frac{-(r-r_{in})^{2} }{8} }  & r>r_{\mathrm{in}} \\0 & r \leq r_{\mathrm{in}}.\end{array}\right.
    \label{equ_35}.
\end{equation}
Among these, \(r_{in}\) is the innermost radius of the accretion disk. In subsequent discussions, three types of inner radii will be considered: the innermost stable circular orbit where \(r_{in}=r_{isco}\) , the photon ring radius where \(r_{in}=r_{ph}\), and the event horizon radius where \(r_{in}=r_{h}\)
\par
First, we consider the inner radius of the thin disk to be the radius of the Innermost Stable Circular Orbit (ISCO), \(r_{isco}\). In this model, there is no emission inside the ISCO radius. For spherically symmetric spacetime, \( r_{isco} \) is given by the following expression
\begin{equation}
    r_{isco}=\frac{3f(r_{isco}){f}'(r_{isco})}{2{f}'(r_{isco})^{2}-f(r_{isco}){f}''(r_{isco})}.
    \label{equ_36}
\end{equation}

\par
For an EMD black hole with \( q_{e} = 0.5 \) and \( M = 1 \), the radius of the \(ISCO\) is approximately \( r_{isco} \simeq 5.61 \). Using Eqs.(\ref{equ_34} - \ref{equ_35}), we can plot the total emitted intensity \( I_{\mathrm{emi}}^{\mathrm{d}} \) as a function of radius, the total observed intensity \( I_{o} \) as a function of collision parameter, and a two-dimensional image in observer's celestial coordinates. Such as the three images in the first column of Figure 14.
\begin{figure*}[htbp]
  \centering
  \includegraphics[width=5cm,height=4cm]{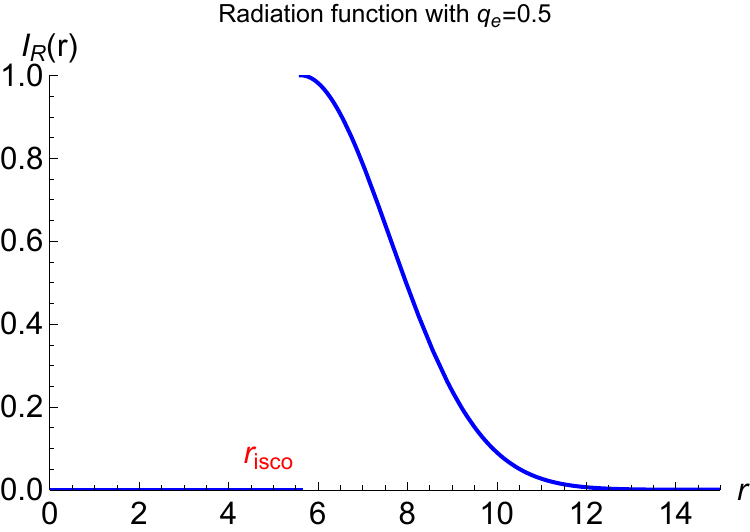}
  \includegraphics[width=5cm,height=4cm]{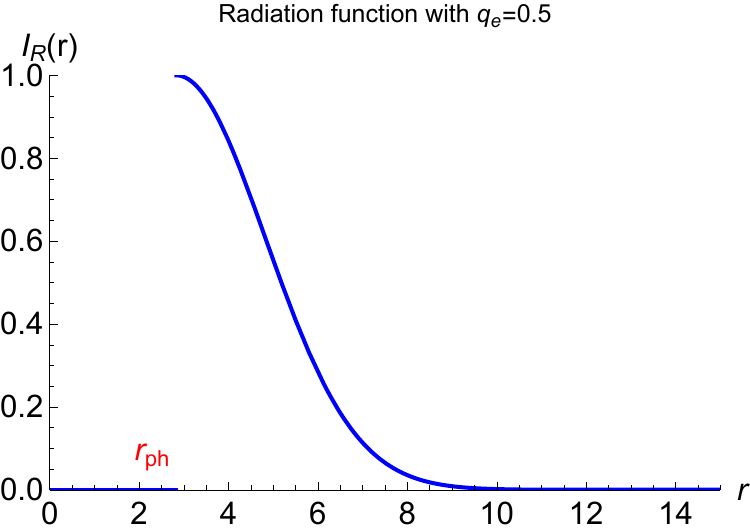}
  \includegraphics[width=5cm,height=4cm]{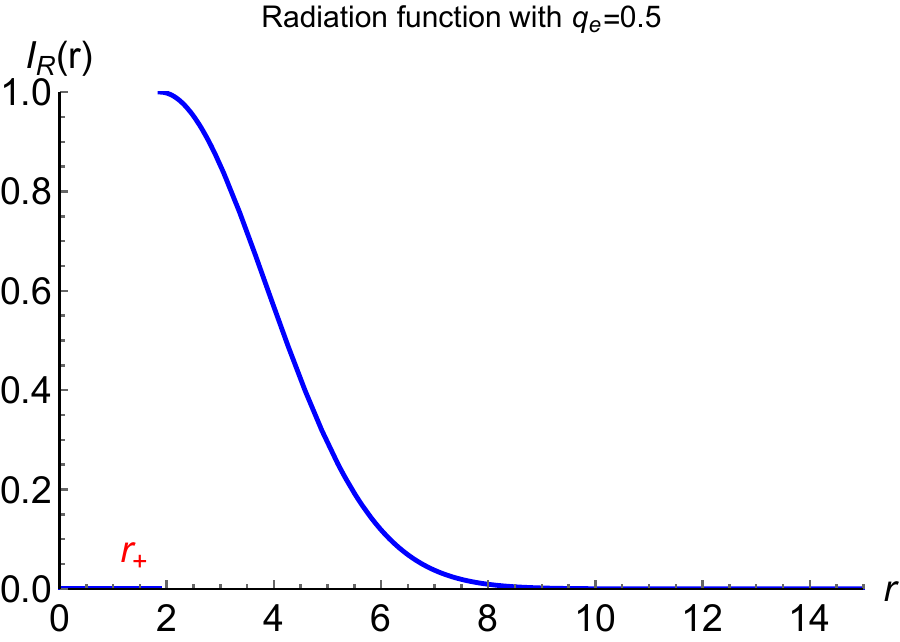}
  \includegraphics[width=5cm,height=4cm]{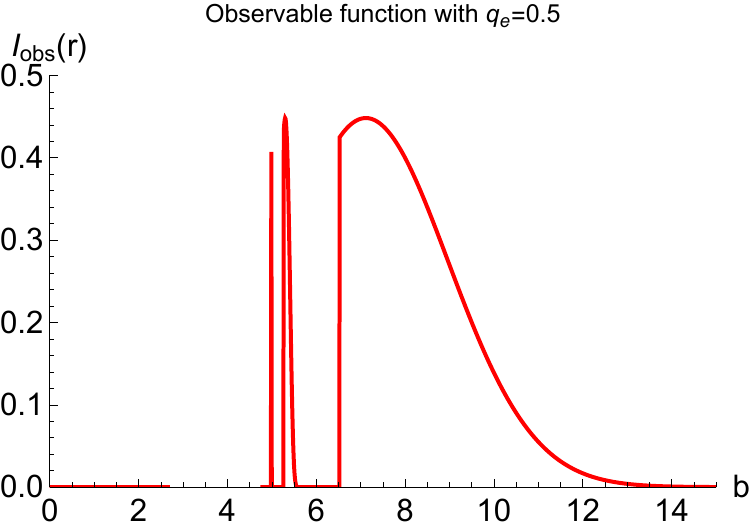}
  \includegraphics[width=5cm,height=4cm]{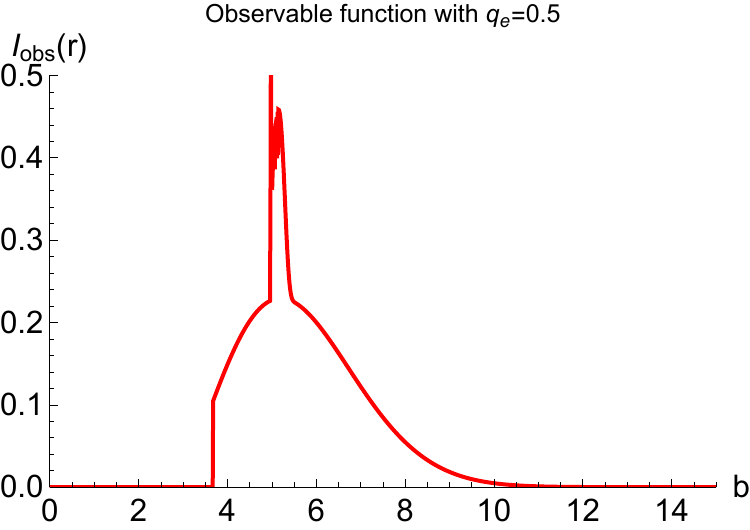}
  \includegraphics[width=5cm,height=4cm]{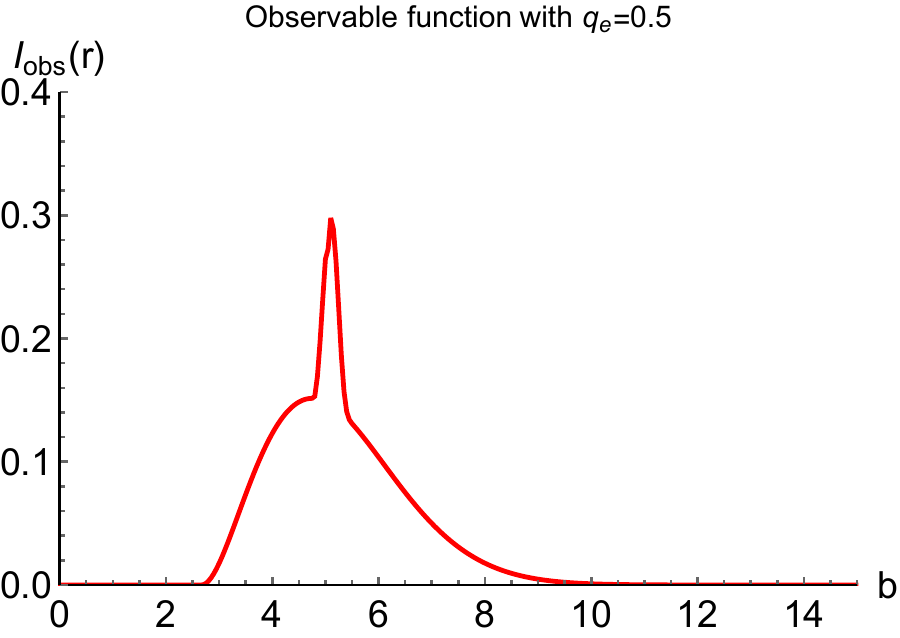}
  \includegraphics[width=5cm,height=4.5cm]{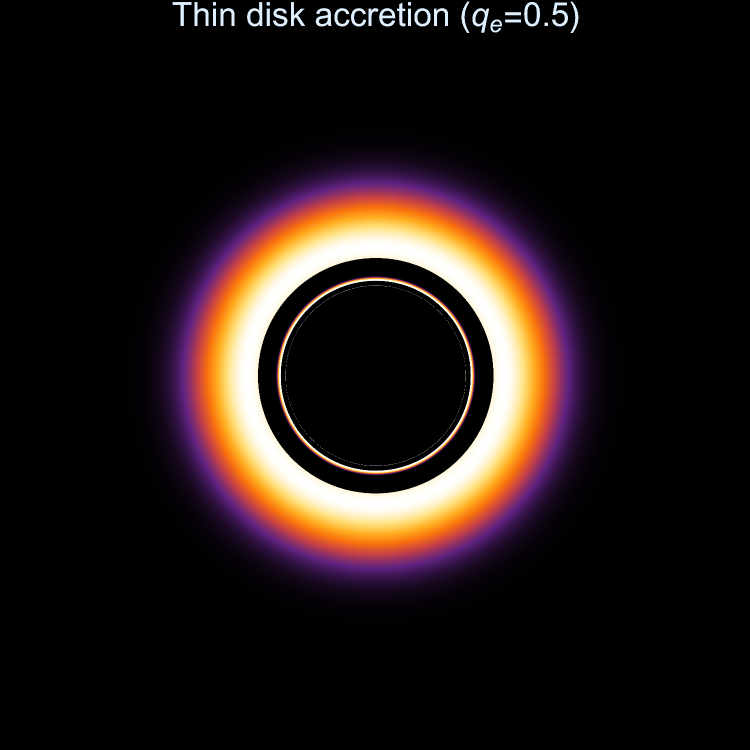}
  \includegraphics[width=5cm,height=4.5cm]{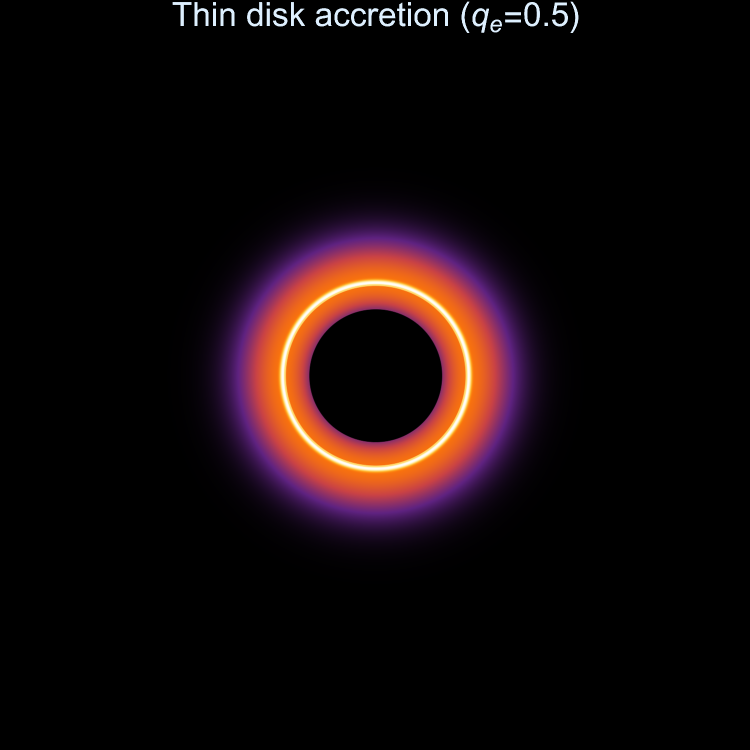}
  \includegraphics[width=5cm,height=4.5cm]{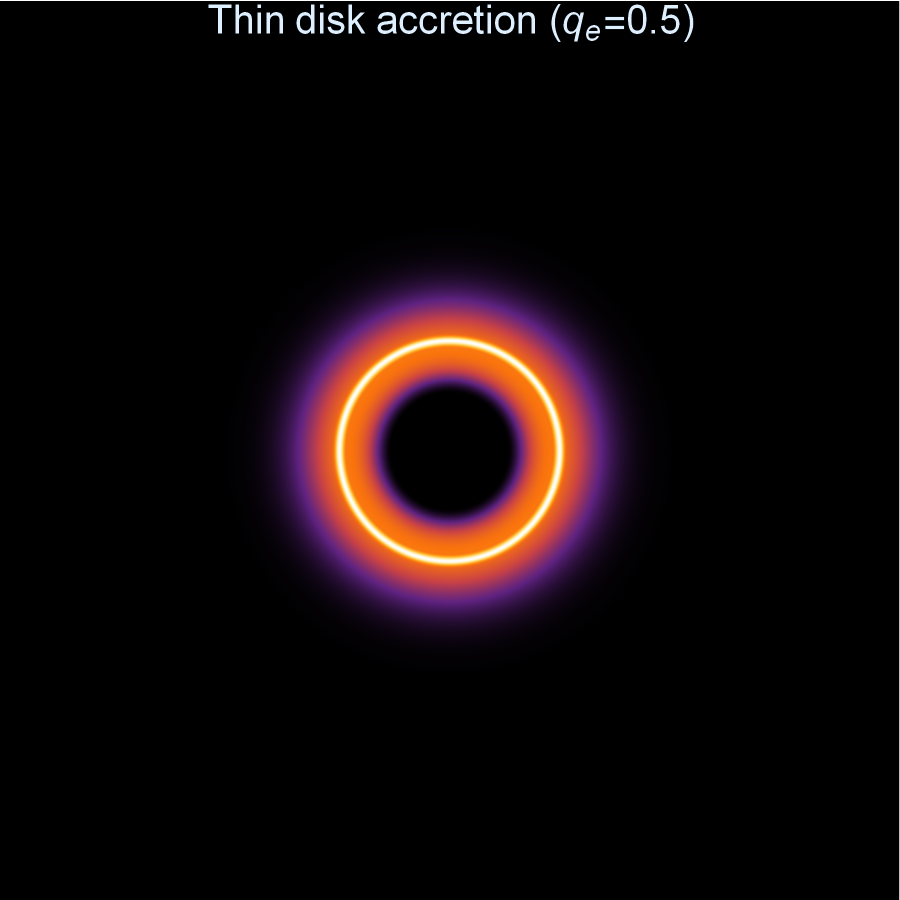}
  \caption {Shadow and ring images of an EMD black hole with \( q_{e} = 0.5 \) under the optical thin and geometrically thin accretion disk model.}\label{fig:14}
\end{figure*}

\par
In this emission model, the total emitted intensity has no radiation within the radius \( r_{isco} \), reaching its peak at \( r_{isco} \simeq 5.61 \), and then decreases with increasing radius, as shown in the first image of the first column in Figure 14. The second image shows the relationship between the observed total intensity and the collision parameter \( b \). The image exhibits three peaks, with the innermost representing the photon ring, followed by the lensed ring, and finally the direct emission. The peak corresponding to the photon ring is very narrow, indicating that the contribution of the photon ring to the total observed intensity is minimal. In contrast, the peak and area corresponding to the lensed ring are larger than those of the photon ring, resulting in a slightly greater contribution to the total observed intensity. The third peak, which is the broadest, indicates that the direct emission contributes the most to the total observed intensity. The bottom row displays the two-dimensional image of this model, showing three rings. The outermost ring corresponds to direct emission, characterized by its maximum thickness and brightness, thereby making the largest contribution to the total observed intensity. The innermost ring represents the photon ring, which is extremely thin and faint, making it nearly invisible in the image. The middle ring corresponds to the lensed ring, which is thicker and brighter compared to the photon ring.
\begin{figure*}[htbp]
  \centering
  \includegraphics[width=5cm,height=4cm]{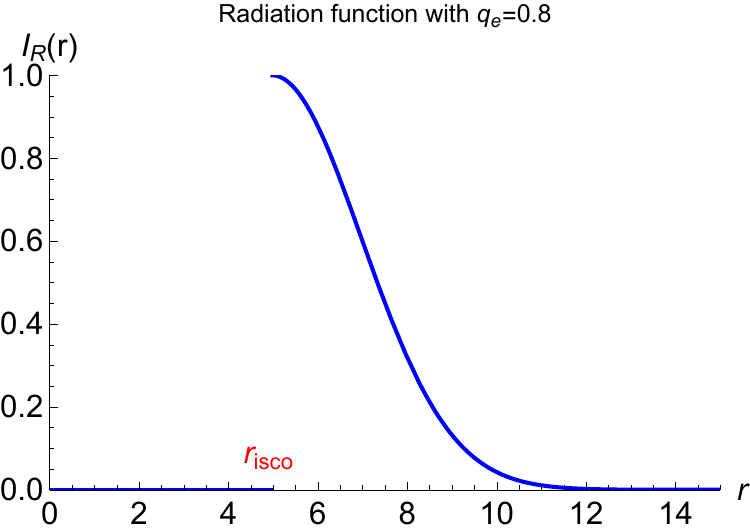}
  \includegraphics[width=5cm,height=4cm]{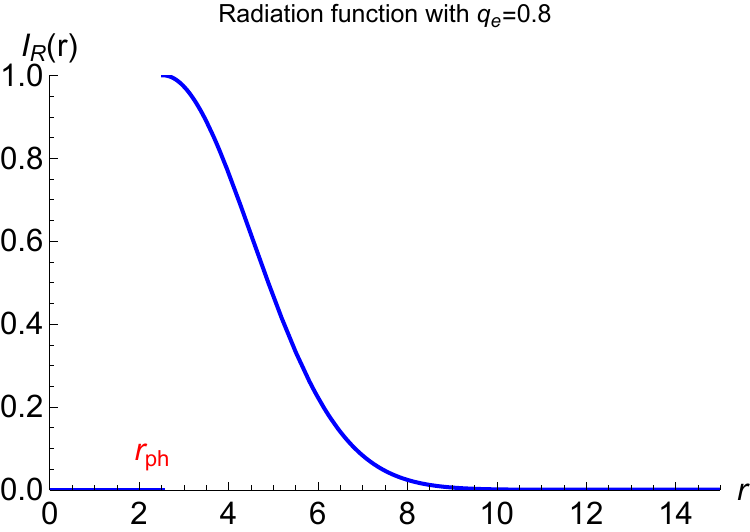}
  \includegraphics[width=5cm,height=4cm]{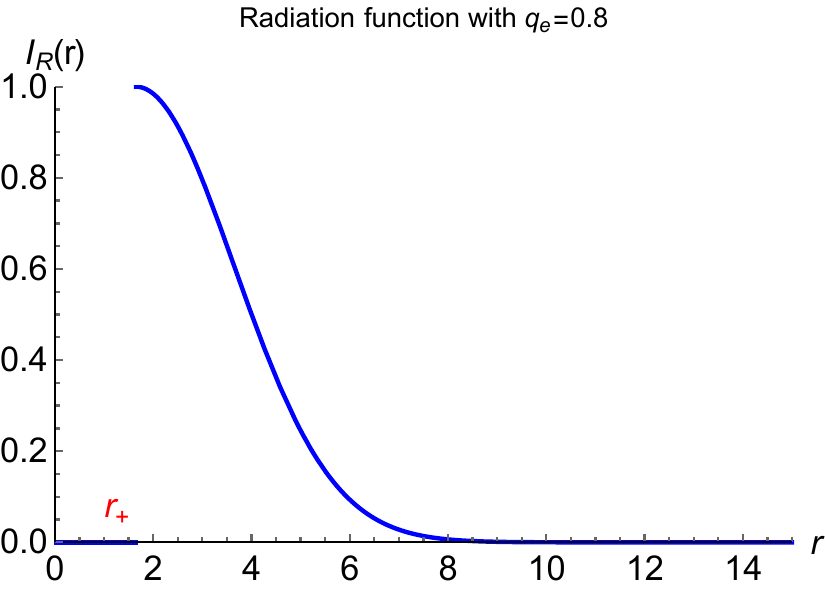}
  \includegraphics[width=5cm,height=4cm]{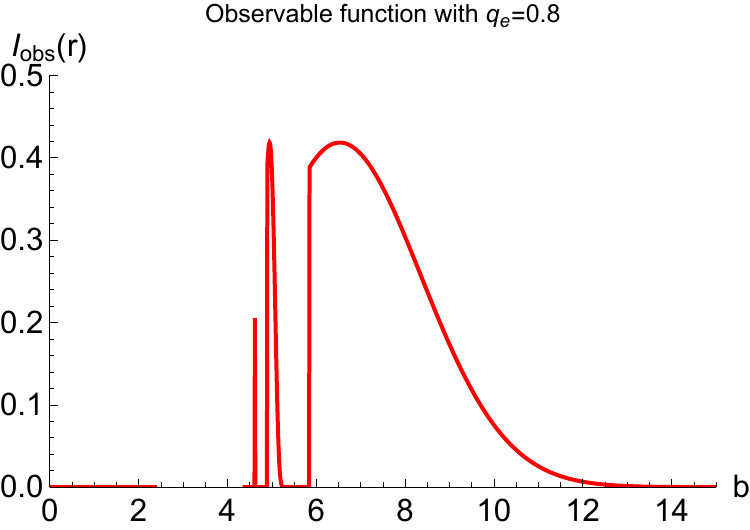}
  \includegraphics[width=5cm,height=4cm]{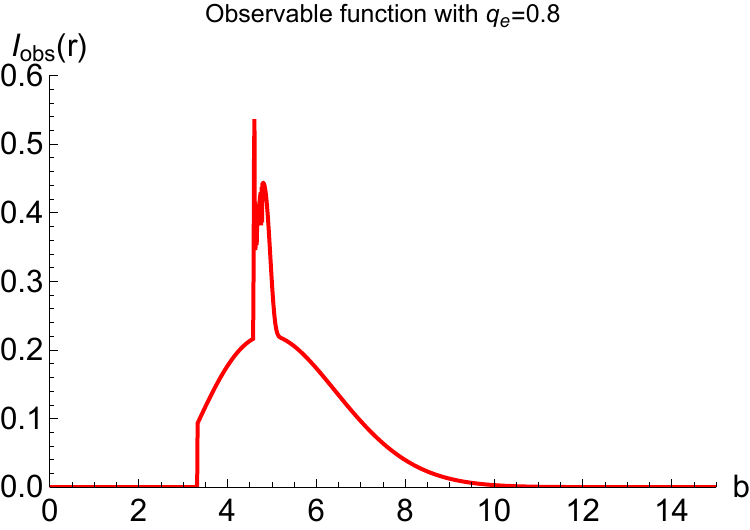}
  \includegraphics[width=5cm,height=4cm]{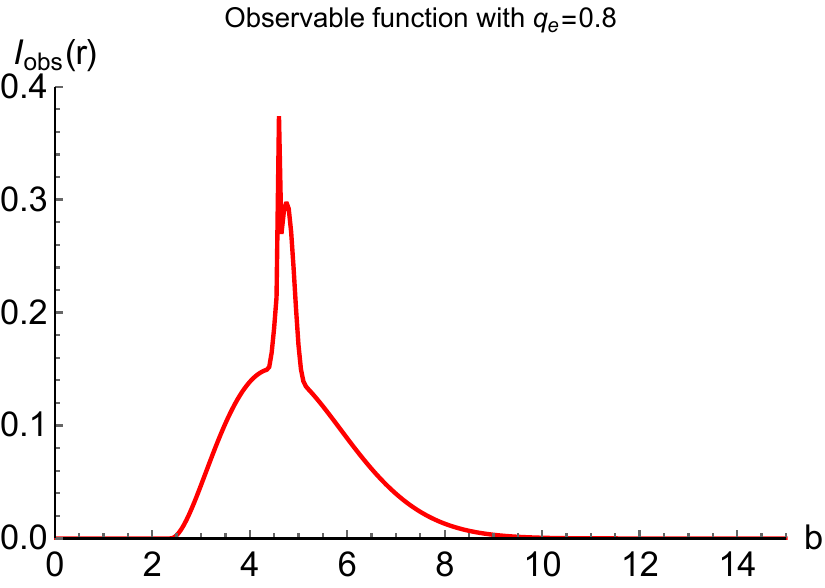}
  \includegraphics[width=5cm,height=4.5cm]{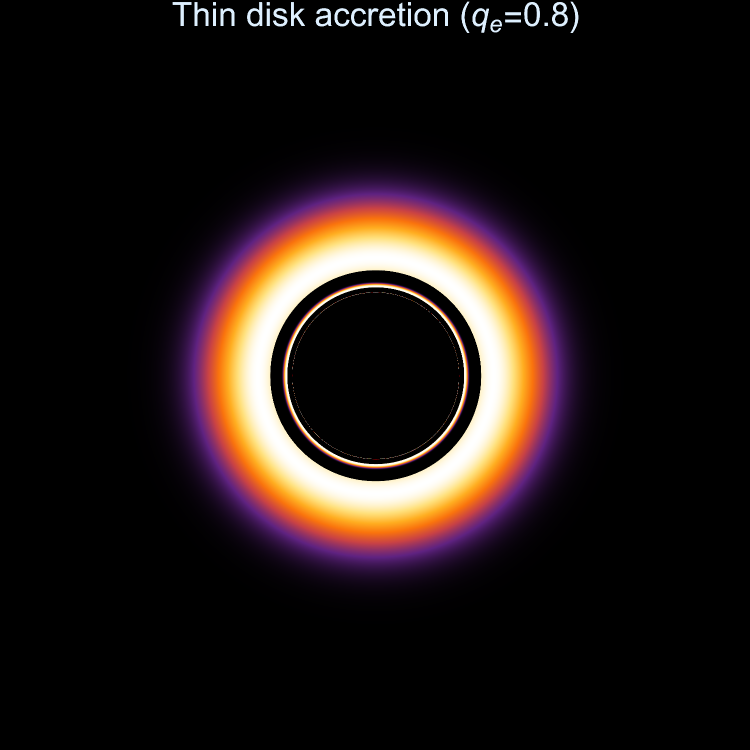}
  \includegraphics[width=5cm,height=4.5cm]{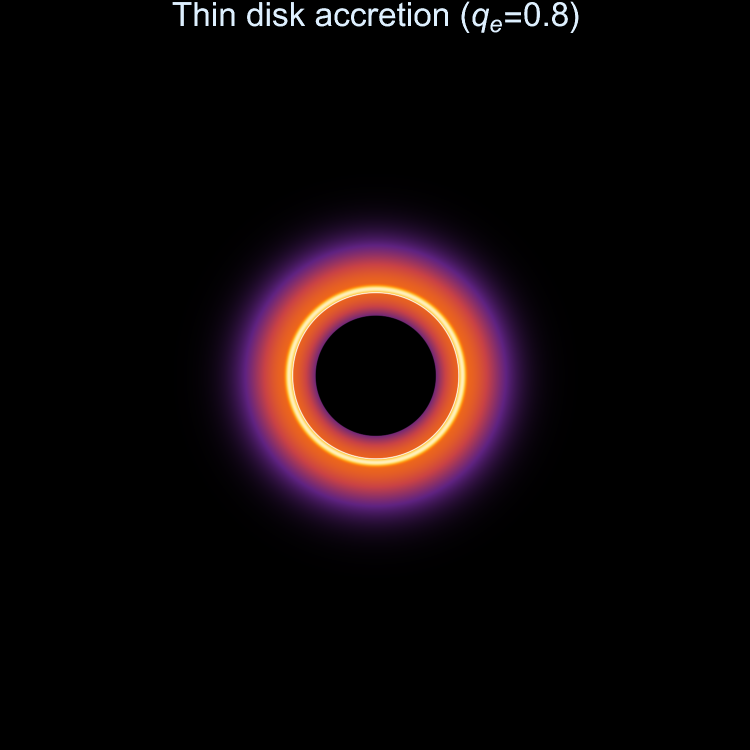}
  \includegraphics[width=5cm,height=4.5cm]{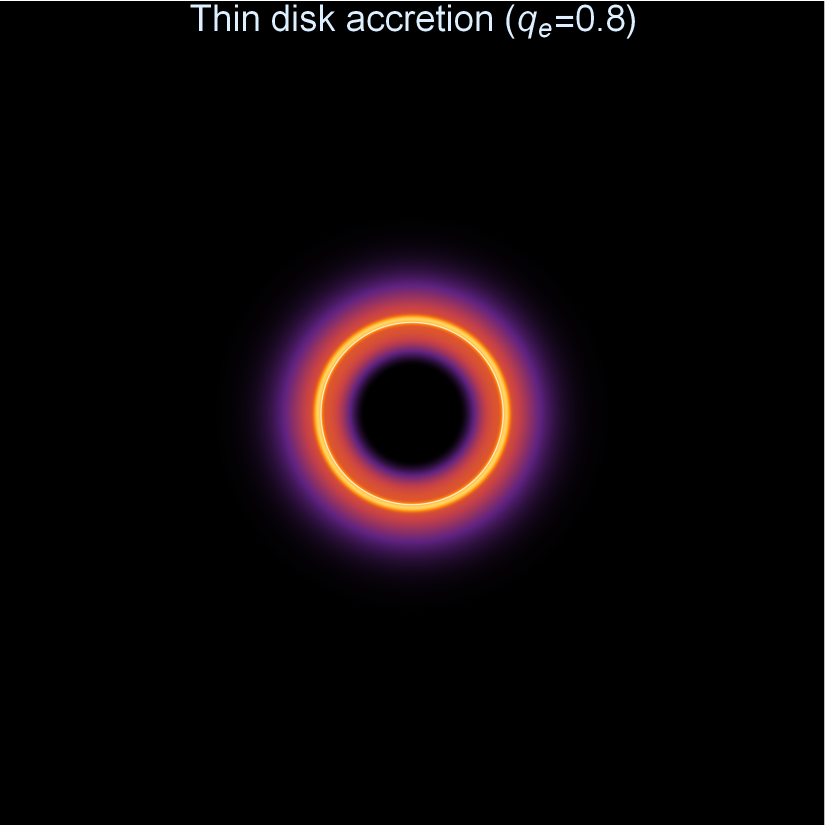}
  \caption {The shadow and ring images of EMD black holes with \( q_{e} = 0.8 \) under the optical thin and geometric thin models of different accretion disk inner radii.}\label{fig:15}
\end{figure*}

\par
Next, we consider the inner radius of the thin accretion disk located at the photon ring radius \( r_{ph} \). In this case, there is no radiation inside the photon ring. The relationship between \( I_{\mathrm{emi}}^{\mathrm{d}} \) and radius in this emission model, the relationship between \( I_{o} \) and collision parameter \( b \), and the two-dimensional image of the black hole are depicted in the second column of Figure 14. From the image, it is evident that the direct emission overlaps with the bright ring formed by the photon ring and the lensing ring. Direct emission begins at \(b=3.67\). The narrower peak represents the photon ring, while the broader but shorter peak corresponds to the lensing ring. Although the photon and lensing rings have higher peak brightness, their range is narrower compared to the direct emission, resulting in the observer primarily experiencing brightness dominated by direct emission.

\par
For the third case, the inner radius will be considered to be at the event horizon radius \(r_{h}\), with its radiation characteristics as shown in the third column of Figure 14. It can be seen that the total emission intensity peaks at the event horizon \(r_{h}\). In this model, the width of the lensing ring is still larger than that of the photon ring, indicating that the lensing ring contributes more to the radiation than the photon ring. Outside the lensing ring region, the direct emission dominates. From the bottom images in the three columns, it can be seen that the photon ring is also included within the lensing ring. Figure 15 shows the observational features of EMD black holes surrounded by thin accretion disk models for different dilaton charges. It can be observed that for EMD black holes, as \( q_{e} \) increases, the range of impact parameters \( b \) corresponding to direct emission, lensing ring, and photon ring gradually decreases.

\subsection{Optical thick accretion model}
\label{sec:3-4}
\par
\begin{figure*}[htbp]
  \centering
  \includegraphics[width=5cm,height=4.5cm]{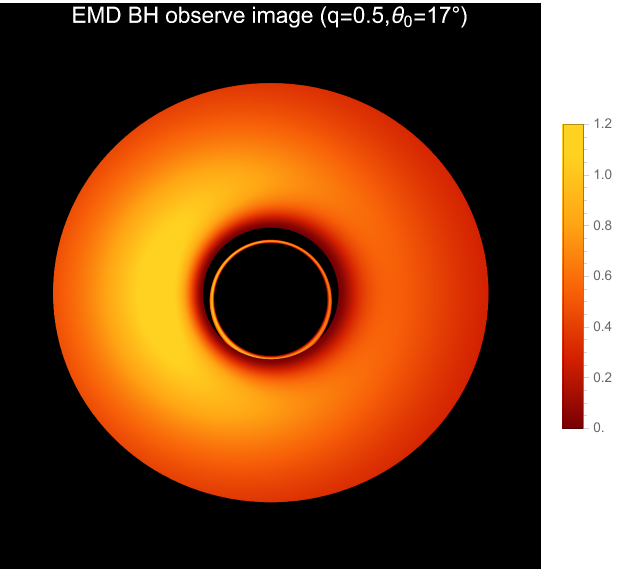}
  \includegraphics[width=5cm,height=4.5cm]{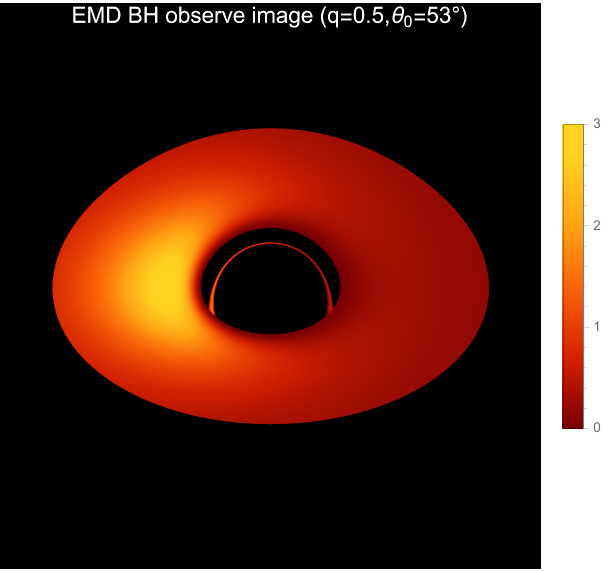}
  \includegraphics[width=5cm,height=4.5cm]{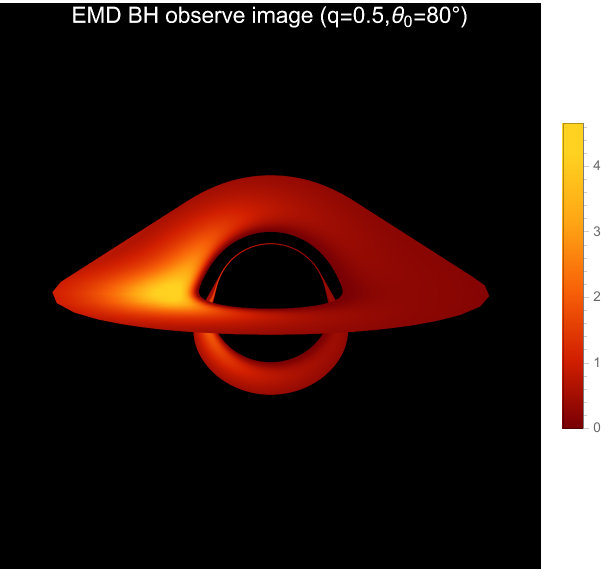}
  \includegraphics[width=5cm,height=4.5cm]{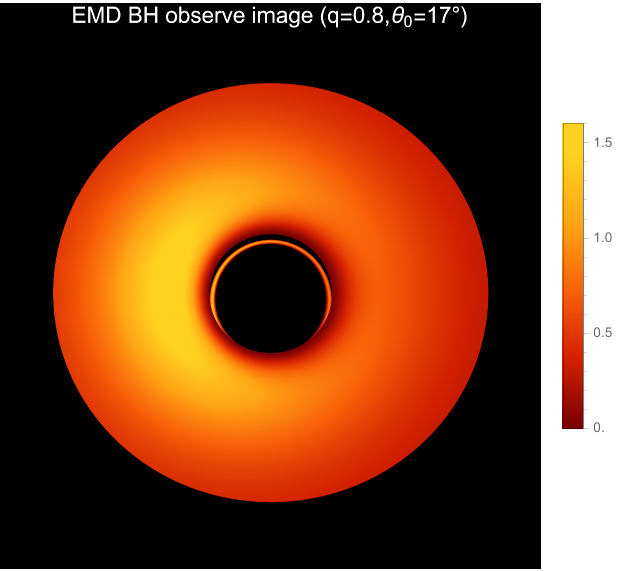}
  \includegraphics[width=5cm,height=4.5cm]{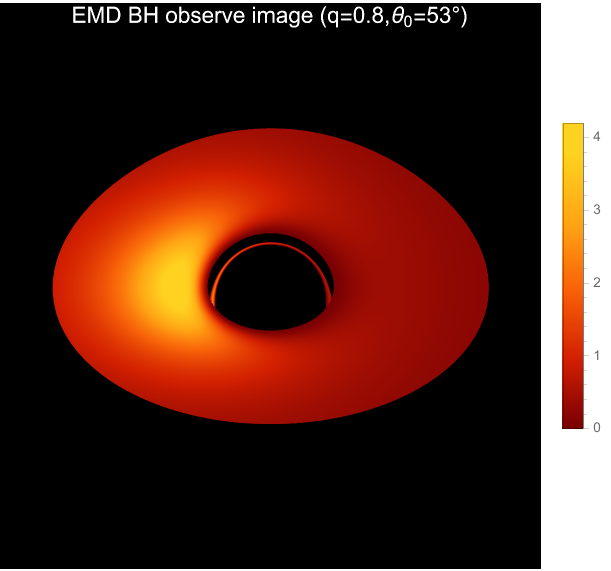}
  \includegraphics[width=5cm,height=4.5cm]{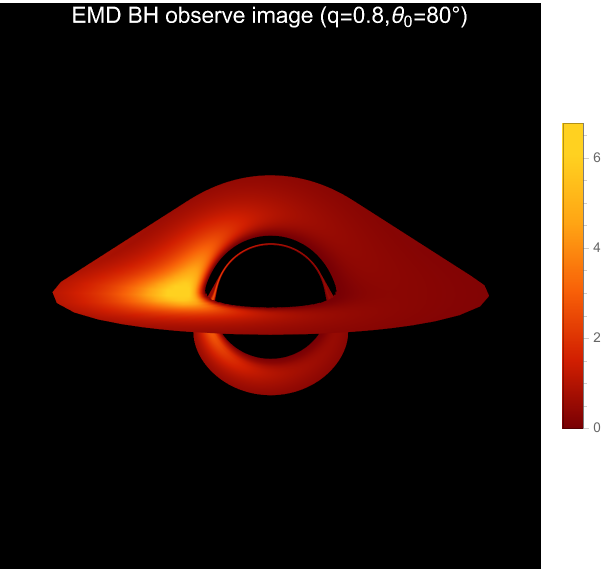}
  \caption {Black hole images observed at different observational inclinations. There is a region of enhanced light intensity on the left side of the direct and secondary images at different observing tilts. The top row is for \(q_{e}=0.5\) and the bottom row is for \(q_{e}=0.8\). There is not much difference in the shapes of the two, but there are some differences in the inner radius and the intensity of the radiation.}\label{fig:16}
\end{figure*}

The previously considered models are all optically thin, therefore, the following section will discuss the observational characteristics of images of geometrically thin optically thick accretion disk models. According to reference \cite{41}, for a thin accretion disk, the radiation intensity emitted at a specific radial position satisfies
\begin{equation}
    F=-\frac{\dot{M}}{4 \pi \sqrt{-\mathrm{g}}} \frac{\Omega, \mathrm{r}}{(E-\Omega L)^{2}} \int_{r_{\mathrm{in}}}^{r}(E-\Omega L) L_{, \mathrm{r}} \mathrm{d}r,
    \label{equ-39}
\end{equation}
where \(\dot{M}\) is the mass accretion rate, \(g\) is the metric determinant, \(r_{in}\) denotes the inner edge of the accretion disk. \(E\), \(\Omega\), and \(L\) represent the energy, angular velocity, and angular momentum of particles on circular orbits, respectively. For a static spherically symmetric black hole, the metric can be written as \(\mathrm{d}s^{2}=g_{tt}\mathrm{d}t^{2}+g_{rr}\mathrm{d}r^{2}+g_{\theta \theta }\mathrm{d}\theta ^{2}+g_{\phi \phi }\mathrm{d}\phi^{2}\). Then \(E\), \(\Omega\) and \(L\) can be expressed as
\begin{equation}
    E=-\frac{g_{tt}}{\sqrt{-g_{tt}-g_{\phi \phi }\Omega ^{2}} },
     \label{equ-40}
\end{equation}
\begin{equation}
    L=\frac{g_{\phi \phi}\Omega}{\sqrt{-g_{tt}-g_{\phi \phi }\Omega ^{2}} },
     \label{equ-41}
\end{equation}
\begin{equation}
    \Omega=\frac{\mathrm{d}\phi}{\mathrm{d}t} =\sqrt{-\frac{g_{tt,r}}{g_{\phi\phi,r }} }.
     \label{equ-42}
\end{equation}
From the above equation, the radiation flux on the disk can be rewtitten as 
\begin{equation}
 F(r) = \frac{\dot{M}\mathcal{H} \sqrt{\frac{M^{4}}{\mathcal{B}}} \mathcal{B}^{2} \mathcal{G}}{2 \pi r \sqrt{r^{2}}  \mathcal{D}   \mathcal{C}^{2}} \int_{r_{i n}}^{r} \frac{8 M^{3} r^{4}   \mathcal{D}^{2}  \mathcal{C}  \mathcal{E}}{ \mathcal{A} \sqrt{\frac{M^{4}}{\mathcal{B}}} \mathcal{B}^{4} \mathcal{F}^{2}} \mathrm{dr},
     \label{equ-43}
\end{equation}
where
\begin{equation}
    \nonumber
    \mathcal{A} = \sqrt{\frac{q^4}{M^2 r^2} + 4},
\end{equation}
\begin{equation}
    \nonumber
    \mathcal{B} = M q^4 r A + 4 M^2 q^2 r^2 + 2 M^3 r^3 A + q^6,
\end{equation}
\begin{equation}
    \nonumber
    \mathcal{C}= -12 M^4 r^2 + M q^4 r A - 2 M^2 (q^4 - 2 q^2 r^2) + 2 M^3 r (r^2 - q^2) A + q^6,
\end{equation}
\begin{equation}
    \nonumber
    \mathcal{D}=4 M^2 r^2 + q^4,
\end{equation}
\begin{align}
\nonumber
    & \mathcal{E}= -24 M^8 r^6 + M^6 (12 q^2 r^6 - 38 q^4 r^4) + M^4 (19 q^6 r^4 - 16 q^8 r^2)  \\ \nonumber
    & + M q^{12} r A - 2 M^2 q^{10} (q^2 - 4 r^2) + 2 M^7 r^5 (r^2 - 9 q^2) A \\ \nonumber
    & + 3 M^5 q^4 r^3 (3 r^2 - 4 q^2) A + 2 M^3 q^8 r (3 r^2 - q^2) A + q^{14},
\end{align}
\begin{equation}
    \nonumber
    \mathcal{F}= q^4 r (r - 2 M A) + 2 M^2 r^3 (2 r - 3 M A) + 8 M^2 q^2 r^2 + 2 q^6,
\end{equation}
\begin{equation}
    \nonumber
    \mathcal{G}=  \frac{2 (3 M^2 r^2 + q^4)}{M r^3 A} - \frac{2 q^2}{r^2} - 1,
\end{equation}
\begin{equation}
    \nonumber
    \mathcal{H}=-M q^{2} r A+6 M^{2} r^{2}+q^{4}.
\end{equation}

\par
Due to the varying gravitational influence of the black hole at different radii on the accretion disk, and also because of the rotational motion of the disk leading to varying velocities and directions at different positions, the calculation of observed radiation flux typically needs to account for redshift effects. These redshift effects primarily include gravitational redshift and Doppler effects. The expression for the redshift factor is
\begin{equation}
    1+z=\frac{E_{em}}{E_{obs}},
     \label{equ-44}
\end{equation}
where \(E_{em}\) is the energy of the photon emitted, represented as the dot product of the photon's four-momentum \(p\) and the four-velocity \(\mu\) of the emitting particle orbiting the black hole, which can be exprssed  as 
\begin{equation}
    E_{em}=p_{t}u^{t}+p_{\phi}u^{\phi}=p_{t}u^{t}(1+\Omega \frac{p_{\phi}}{p_{t}} ),
     \label{equ-45}
\end{equation}
\(p_{t}\) and \(p_{\phi}\) represent the photon four-momentum components. For a distant observer, the ratio \(p_{t}/p_{\phi}\) represents the impact parameter of the photon relative to the \(z\)-axis, related to trigonometric functions as follows

\begin{equation}
    \frac{p_{t}}{p_{\phi }} =b\sin\theta_{0}\sin\alpha.
     \label{equ-47}
\end{equation}
The redshift factor is therefore can be written as 
\begin{equation}
    1+z=\frac{E_{em}}{E_{obs}}=\frac{1+b\Omega \cos\beta }{\sqrt{-g_{tt}-\Omega^{2}g_{\phi \phi }} }.
     \label{equ-48}
\end{equation}
The redshift factor for EMD black holes in this case  is
\begin{equation}
    1+z= \frac{1+2b \sqrt{\frac{M^4}{M q^4 r \mathcal{A}+4 M^2 q^2 r^2+2 M^3 r^3 \mathcal{A}+q^6}} \sin\alpha \sin\theta_{0} }{\sqrt{1+\frac{ 2q_{e}^{2}}{r^{2}}-\frac{2(q_{e}^{4}+3M^{2}r^{2})}{\mathcal{A} r^{3}} } },
     \label{equ-49}
\end{equation}
where \(\mathcal{A}\) is defined as before. Therefore, for the observed accretion disk image, considering the redshift effect, we have
\begin{equation}
    F_{obs}=\frac{F}{(1+z)^{4}}.
     \label{equ-50}
\end{equation}
This equation can be used to plot the radiation intensity map of the accretion disk of EMD black holes as shown in Figure 16. From the figure, it can be observed that in all viewing angles, the accretion disk exhibits an enhanced radiation intensity on the left side. This occurs because the rotation of the accretion disk causes the observed light intensity to be affected by the Doppler effect and also by gravitational redshift in the presence of a strong gravitational field. Since the accretion material on the left side of the disk moves toward the observer, while the right side moves away from the observer, the flux observed on the left side is larger. Additionally, with increasing viewing angle, the direct and secondary images gradually separate, with the direct image exhibiting a "hat-like" shape.

\section{\textbf{\textcolor{blue}{Summary And Discussion}}}
\label{sec:4}
\par
This study investigates the optical appearance of EMD black holes under various conditions. Specifically, we derive the effective potential of the EMD black hole and determine the null geodesics of massless particles around it. Using a ray-tracing code, we plot the trajectories of photons around the black hole. It is observed that as the dilaton charge of the EMD black hole increases, both the event horizon radius \(r_{h}\) and the photon sphere radius decrease. By examining the relationship between the observer's inclination angle, orbital radius, and celestial coordinates, we plot the direct and secondary images for different observer inclination angles. The results show that as the observer inclination angle increases, the secondary images, initially nested within the direct images, gradually separate. Additionally, the direct images evolve from circular rings to a shape resembling a hat.

Next, we explore the impact of accretion material on the images of the EMD black hole under different accretion models. In the case of a spherically symmetric, geometrically thin, and optically thin accretion model, we find that, regardless of whether the accretion material falls inward, the size of the black hole shadow and the radius of the photon sphere remain unchanged under different conditions. Therefore, the shape of the black hole shadow and the photon ring is determined solely by the spacetime of the black hole itself and is independent of the accretion material. However, when accretion material falls into the EMD black hole, Doppler effects cause the overall brightness of the black hole image to decrease compared to the static scenario.

When considering a geometrically thin, optically thick accretion disk model, we define the direct emission, lensed ring, and photon ring based on the number of intersections between light trajectories and the accretion disk. By varying the inner radius of the accretion disk, we can adjust the radii of these light rings, thereby producing different black hole images. In these light rings, the direct emission is the dominant component of the black hole brightness, followed by the lensed ring, with the photon ring making a negligible contribution. Finally, by examining the optically thick, geometrically thin accretion disk model, we observe that the observed radiation intensity is influenced by gravitational redshift and Doppler effects. As a result, the radiation intensity is highest on the left side near the black hole, and as the observer's inclination angle increases, the image becomes progressively more asymmetric, with the distribution of radiation intensity varying accordingly.

EMD black holes is constructed through the non-minimal coupling of the dilaton field with the Maxwell field, and its action is given by Eq.(\ref{equ-1}). The parameter \(\alpha\) an take arbitrary values, which in this study is reflected in the dilaton charge \(q\). However, when the dilaton coupling parameter in Eq.(\ref{equ-1}) is fixed to a specific value, namely \(\alpha=1\), the renowned Gibbons-Maeda dilaton spacetime can be derived\cite{Aydogdu2006}
\begin{align}
    \nonumber
    d s^{2}= \nonumber
    &  \frac{\left(r-r_{+}\right)\left(r-r_{-}\right)}{r^{2}-D^{2}} d t^{2}-\frac{r^{2}-D^{2}}{\left(r-r_{+}\right)\left(r-r_{-}\right)} d r^{2}\\ \nonumber
    &-\left(r^{2}-D^{2}\right)\left(d \theta^{2}+\sin ^{2} \theta d \phi^{2}\right).
       \label{equ_51}
\end{align}
Here, \(D=\frac{P^{2}-Q^{2}}{2M}\), where the parameters \(P\) and \(Q\) denote the magnetic charge and electric charge of the black hole. When \(P=0\), he Garfinkle-Horowitz-Strominger dilaton spacetime is obtained. Both EMD black holes and GHSd black holes arise from the non-minimal coupling of the dilaton field with the Maxwell field, suggesting that there may be many interesting relationships between them. When the hair parameters of EMD black holes and GHSd black holes take the same values, their horizon radii differ, but their shadow radii remain identical. For instance, when \(q_{e}=0.5\) and \(Q=0.5\), the shadow radii of both are \(4.9732M\). Furthermore, the shadow images of these two black holes will also exhibit some differences.
\begin{figure*}[htbp]
\centering
\includegraphics[scale=0.6]{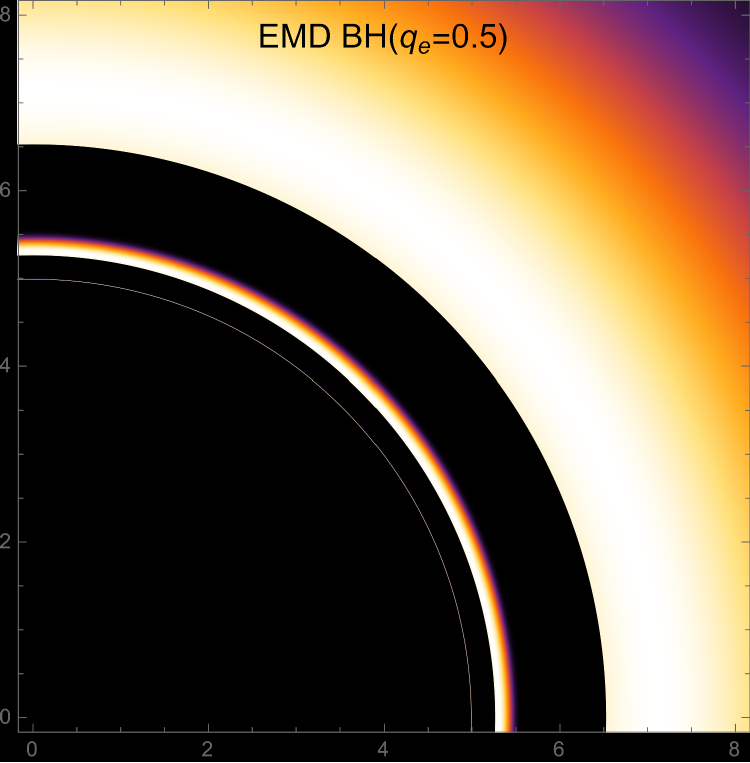}
\includegraphics[scale=0.6]{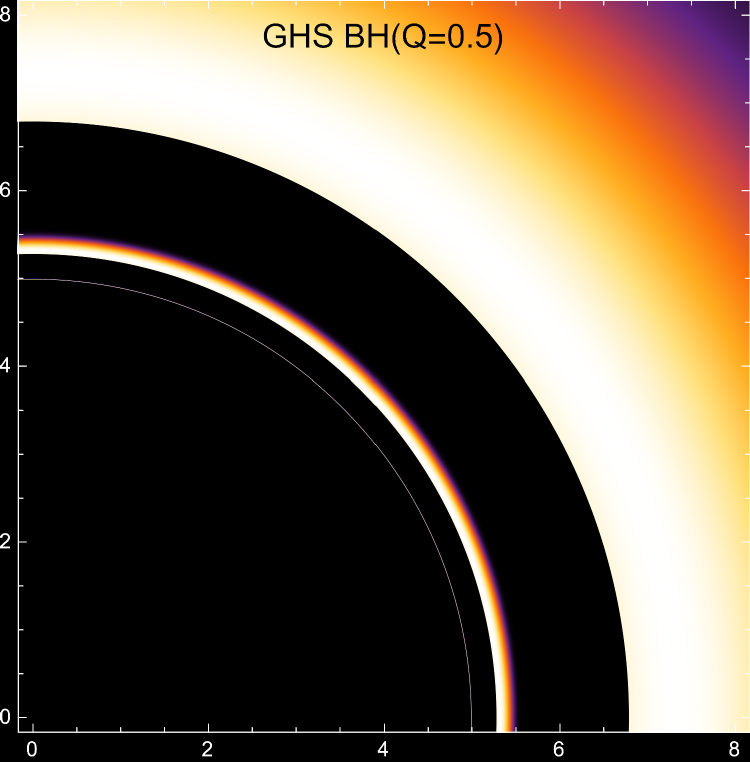}
\caption {The image is the observed simulation image under the optically thin geometric thin accretion model when the inner diameter of the accretion disk is \(R_{isco}\). The left image is the EMD black hole image when \(q_{e}=0.5\), and the right image is the GHS black hole image when \(Q=0.5\).}\label{fig:17}
\end{figure*}
To observe the differences between the images of the EMD black hole and the GHSd black hole, we performed a local magnification of the observational images obtained using the method described in Section 3.3, highlighting the details of the photon ring and lensing ring structures, as shown in Figure 17. Through comparative analysis, it is observed that the photon ring and lensing ring structures of the EMD black hole and the GHSd black hole exhibit a high degree of similarity in overall morphology and radiation distribution. However, it is important to note that there is a significant difference in the gap between the lensing ring and the direct radiation zone, with the gap width noticeably increased. The underlying causes of this phenomenon will be the central focus of our future research.

\Acknowledgements{This work is supported  by the National
Natural Science Foundation of China (Grants No. 11675140, No. 11705005, and No. 12375043), and Innovation and Development Joint  Foundation of Chongqing Natural Science  Foundation (Grant No. CSTB2022NSCQ-LZX0021) and Basic Research Project of Science and Technology Committee of Chongqing (Grant No. CSTB2023NSCQ-MSX0324), and Fund Project of Chongqing Normal University (Grant Number: 24XLB033).}

\InterestConflict{The authors declare that they have no conflict of interest.}








\end{multicols}
\end{document}